\documentclass[a4paper,fleqn]{cas-dc}

\usepackage[authoryear]{natbib}
\begin{document}
\let\WriteBookmarks\relax
\def\floatpagepagefraction{1}
\def\textpagefraction{.001}
\shorttitle{Application of an Upwind Integration Method to Plane Parallel Hall-MHD}
\shortauthors{Georgios Chouliaras and K.N. Gourgouliatos}

\title [mode = title]{Application of an Upwind Integration Method to Plane Parallel Hall-MHD}                       

\author[1,2]{Georgios Chouliaras}[type=editor,
                        auid=000,bioid=1]
\cormark[1]
\ead{gc205@st-andrews.ac.uk}

\credit{Conceptualization of this study, Methodology, Software}

\address[1]{University of Patras, Department of Physics, 26504,  Patras, Greece}
\address[2]{University of St.Andrews, School of Mathematics and Statistics, St Andrews KY 16 9SS, UK}

\author[1]{Konstantinos N. Gourgouliatos}
\cormark[1]
\ead{kngourg@upatras.gr}

\credit{Conceptualization of this study, Methodology, Software}

\cortext[cor1]{Corresponding author}

\begin{abstract}
 \textbf{Aims:} We study the impact of an Upwind scheme on the numerical convergence of simulations of the Hall and Ohmic effect in neutron stars crusts. While simulations of these effects have explored a variety of geometries and wide ranges of physical parameters, they are limited to relatively low values of the Hall parameter, playing the role of the magnetic Reynolds number, which should be not exceed a few hundred for numerical convergence. \\
 \textbf{Methods}: We study the evolution of the magnetic field in a plane-parallel Cartesian geometry. We discretise the induction equation using a finite difference scheme and then integrate it via the Euler forward method. Two different approaches are used for the integration of the advective terms appearing in the equation: a Forward Time and Central in Space (FTCS) and an Upwind scheme. We compare them in terms of accuracy and performance. We explore the impact of the Upwind method on convergence according to the ratio of planar to vertical field and the Hall parameter.\\
\textbf{Results}: In the limit of a low strength planar field the use of an Upwind scheme provides a vast improvement leading to the convergence of simulations where the Hall parameter is 2 orders of magnitude higher than that of the FTCS. Upwind is still better if the planar field is stronger, yet, the difference of the maximum value of the Hall parameter reached is within a factor of 10 or a few. Moreover, we notice if the schemes diverge their behaviour is very different, with FTCS producing infinite energy, while the Upwind scheme only temporarily increasing the overall magnetic field energy. \\
\textbf{Conclusions}: Overall, the Upwind scheme enhances the efficiency of the simulations allowing the exploration of environments with higher value of  electric conductivity getting us closer than before to realistic environmental conditions of magnetars.
\end{abstract}


\maketitle

\section{Introduction}

\label{intro}
Several decades of observations of strongly magnetised neutron stars (NSs) have revealed violent phenomena such as outbursts and flares in magnetars, which is related to magnetic field evolution \citep{1999ApJ...519L.151M, 2015RPPh...78k6901T, 2017ARA&A..55..261K, 2018IAUS..337..326Z, 2018MNRAS.474..961C}. Furthermore, their thermal evolution indicates that their magnetic field is powering, at least part of, their thermal evolution through the conversion of magnetic energy into heat \citep{1990A&A...229..133H, 1997A&A...321..685S, 2007PhRvL..98g1101P}. Even when including normal rotational powered pulsars, their evolution in the $P-\dot{P}$ diagram and braking indices have been interpreted to hint magnetic field evolution \citep{1988MNRAS.234P..57B, 1994A&AT....4..235U, 1997MNRAS.292..167U, 2007Ap&SS.309..245G, 2014MNRAS.443.1891G, 2015MNRAS.446.1121G, 2015AN....336..831I, 2015AN....336..861P, 2017MNRAS.467.3493J, 2019LRCA....5....3P, 2020MNRAS.499.2826I}. While these effects address the question of the long term evolution, they also provide the necessary triggers for short term events due to physical instabilities leading to explosive events \citep{2015MNRAS.453L..93G}. 

Even before the abundance of magnetar data, magnetic field evolution in the neutron star has been theoretically formulated and attributed to the Hall effect and Ohmic decay in the crust and ambipolar diffusion in the core and deeper crust \citep{1983MNRAS.204.1025B, 1988MNRAS.233..875J, 1992ApJ...395..250G}. Ambipolar diffusion and Ohmic decay are dissipative processes, whereas the Hall effect is conservative. Therefore, its simulation is more complex and prone to physical and numerical instabilities due to the formation of sharp discontinuities and current sheets that are not formally dissipated unless some other dissipate effect is accounted for \citep{2000PhRvE..61.4422V, 2016MNRAS.463.3381G, 2017AstL...43..624K, 2019AN....340..475K}. 

Despite these hurdles, numerical simulations of the Hall effect have been successfully implemented in a spherical geometry under axisymmetry \citep{2002MNRAS.337..216H, 2004MNRAS.347.1273H, 2009A&A...496..207P, 2012CoPhC.183.2042V, 2013MNRAS.434..123V, 2014MNRAS.438.1618G}. Indeed, the simulation of the Hall effect in association with Ohmic decay leads to a manageable Hall parameter (the equivalent of Reynolds number in Hall MHD). However, if the induction equation contains only the Hall term, application of Euler schemes fails to simulate correctly the propagation of the information within the numerical grid and this leads to the instability of the scheme \citep{2003MNRAS.344.1210F}. 

Euler method is the simplest way of numerical integration and straightforward to apply. Realistic neutron stars have high, but finite, conductivity and the actual evolution can be simulated using such schemes, provided the conductivity is low enough so that the ratio of the magnitude of the Hall term over the Ohmic is 10-100  \citep{2014MNRAS.438.1618G}. More sophisticated techniques have been used, such as Godunov shock capturing schemes \citep{2012CoPhC.183.2042V, 2013MNRAS.434..123V}. These approaches have permitted the numerical integration of the Hall-induction equation for higher values of the Hall parameter. Other works have relied on spectral methods for the numerical integration of the equations \citep{2002MNRAS.337..216H, 2004MNRAS.347.1273H,2007A&A...470..303P, 2009A&A...508L..39W,2009ApJ...701..236C, 2010JPlPh..76..117W}. Three dimensional studies, so far, have used mixed approaches, combining spectral and finite difference methods while using Crank-Nicolson and Adams-Bashforth schemes for time integration \citep{2015PhRvL.114s1101W, 2016PNAS..113.3944G}.
Each method has its own strengths and weaknesses and the appropriate choice depends on the nature of the phenomenon the study focuses on. Spectral methods are faster and better suited for spherical systems, naturally matching the geometry of the crust, yet, their implementation on a crust whose density varies over several orders of magnitude may be complicated. Finite difference methods are slower, yet, resolve more efficiently discontinuities. Mixed methods, i.e. spectral in the angular and finite difference in the radial direction remove the issues of density stratification, but may not resolve that well any shocks or strong currents in the angular direction. A Cartesian 3-D code generated through the platform Simflowny has lead to the solution of the generalised induction equation using high order numerical schemes for the time and spatial discretization \citep{2019CoPhC.237..168V}. Numerical integration using the Pencil code has been used for the study of turbulence due to Hall-MHD \citep{2020ApJ...901...18B}. Still in Cartesian geometry, integration of the Hall-MHD equation and exploration of instabilities has been performed using both finite difference and spectral methods \citep{2015MNRAS.453L..93G, 2016MNRAS.463.3381G}. In these works the maximum value of the Hall paremeter reaches was in the range of $\sim 200$.

Overall, these simulations, while they have revealed several important properties of strongly magnetised neutron stars, they have also demonstrated the limitations of the integration schemes. Indeed, the use of Godunov schemes \citep{2012CoPhC.183.2042V} has allowed numerical convergence for higher values of the Hall parameters. However, the nature of the Hall-MHD equations containing advective terms, suggests that a possible improvement is through the use of a relatively simple Upwind scheme. In this work we explore this scenario, using a simplified plane-parallel cartesian problem, and we demonstrate the conditions leading to an enhancement to numerical integration.

The structure of this paper is as follows: We present the mathematical setup of the problem in \S~\ref{setup}. In \S~\ref{Num_setup} we analyse our strategy and our setup of the model and illustrate how Upwind schemes treats them. We present our simulations and the results in section \S~\ref{Sim}. We conclude in section \S~\ref{Con}.

\section{Problem formulation}
\label{setup}
We approximate the neutron star crust by a Coulomb lattice of fixed ions and electrons that are free to move \citep{1992ApJ...395..250G}. The evolution of the crustal magnetic field is described by Hall-MHD, sometimes referred to as electron-MHD in this context, where electrons are the only species moving. Under this approximation, we can relate the electron motion to the electric current:
\begin{eqnarray}
{\bf v}_{\rm e}= -\frac{\bf j}{{\rm e}n_{\rm e}}=-c\frac{ \nabla \times {\bf B}}{{4 \pi {\rm e}  }n_{\rm e}} 
\label{VEL}
\end{eqnarray}
where ${\rm e}$ is the electron charge, $n_{\rm e}$ the electron number density, $c$ the speed of light, and ${\bf B}$ the magnetic field. Ohm's law reads: 
\begin{eqnarray}
{\bf E} = -\frac{{\bf v}_{\rm e}\times {\bf B}}{c}+\frac{\bf j}{\sigma},
\end{eqnarray}
where $\sigma$ is the electric conductivity. Substituting the above expression for the electric field into the induction equation, while neglecting the displacement current, and by virtue of Amp\`ere's law we obtain the following equation for the magnetic field evolution
\begin{equation}
    \frac{\partial{\mathbf{B}}}{\partial{t}}= -\nabla \times \left[\frac{c}{4\pi n_{\rm e}e}(\nabla \times \mathbf{B})\times \mathbf{B} +\frac{c^2}{4\pi \sigma} \nabla \times \mathbf{B}\right]
    \label{HALL}
\end{equation}
The first term on the right hand of the equation (\ref{HALL}) is associated with the Hall effect whilst the second term with Ohmic dissipation. 

Equation \ref{HALL} allows the definition of two characteristic time-scales of the system, the Hall time-scale $t_H=\frac{4\pi e n_e L^2}{c \|B \|}$ with $\|B\|=\max |{\bf B}| $ is the norm of the magnetic field of the system given by maximum value of the modulus of ${\bf B}$ within the integration domain, and Ohmic decay timescale $t_{Ohm}=\frac{4\pi \sigma L^2}{c^2}$. Both have the same, quadratic, dependence on the length-scale of the system and their ratio is the dimensionless Hall parameter:
\begin{eqnarray}
R_{H}=\frac{t_{Ohm}}{t_{H}}=\frac{\sigma \|B\|}{ecn_{\rm e}}.
\end{eqnarray}
The Hall parameter is a dimensionless tool that compares the Hall effect and the Ohmic decay in a system. From Gauss' law $\nabla \cdot {\bf B} =0 $ the magnetic field has zero-divergence, thus we can express it in terms of two scalar functions. In our approach we assume a plane parallel geometry, where the field has all three components, but all physical quantities depend only on $x,~z$. Subject to these constraints we write the magnetic field as follows
\begin{equation}
    \mathbf{B}= B_y \hat{\mathbf{y}} +\nabla \Psi \times \hat{\mathbf{y}}
    \label{BFIELD}
\end{equation}
Where $B_y(x,z)$ is the magnetic field component along the $y$ axis and $\Psi(x,z)$ is a scalar. We refer to $B_y$ as the vertical field and to the components $B_x,~B_z$ as the planar field. The form of $\Psi$ provides the structure of the magnetic field on $x-z$ plane, with contours of constant $\Psi$ being parallel on the $B_{x}\hat{\mathbf x} +B_{z}\hat{\mathbf z}$ component of the field. 

We further use Amp\`ere's law to evaluate the electric current:
\begin{eqnarray}
{\bf j}=\frac{c}{4\pi}\nabla \times {\bf B}=\frac{c}{4\pi}\left(-\nabla^{2}\Psi \hat{\bf y}+\nabla B_{y}\times \hat{\bf y}\right)\,.
\label{CURRENT}
\end{eqnarray}
By virtue of equations (\ref{VEL}, \ref{CURRENT}) we can further evaluate the electron fluid velocity:
\begin{eqnarray}
    {\bf v}_{\rm e}=\frac{c}{4\pi e n_{\rm e}}\left(\nabla^{2}\Psi \hat{\bf y}-\nabla B_{y}\times \hat{\bf y}\right)\,
    \label{V_ADV}
\end{eqnarray}

We substitute the magnetic field from equation (\ref{BFIELD}) into the magnetic field induction equation (\ref{HALL}) and we obtain two coupled partial differential equations for $\Psi$ and $B_y$
\begin{eqnarray}
    \frac{\partial B_y}{\partial t}&=&\frac{-c}{4\pi e}\bigg\{\left[\nabla\left(\frac{\nabla^2 \Psi}{n_e}\right)\times \hat{\mathbf y}\right]\cdot \nabla\Psi  \nonumber \\ 
    &+&B_y\left(\nabla n_e^{-1}\times \hat{\mathbf y}\right)\cdot \nabla B_y  \bigg\}+\frac{c^2}{4\pi \sigma}\nabla^2 B_y \label{dBy} \\
    \frac{\partial\Psi}{\partial t}&=&\frac{c}{4\pi n_e e}\left(\nabla B_y\times \hat{\mathbf y}\right)\cdot \nabla \Psi+\frac{c^2}{4 \pi \sigma}\nabla^2\Psi \label{dPsi} 
\end{eqnarray}
The above equation encapsulates the evolution of the magnetic field due to the Hall effect and Ohmic decay. Regarding the equation for $B_y$ (\ref{dBy}), the first term on the right hand-side shows how the field-lines of the $x-z$ plane are bent into $B_y$, if the $y-$component of the electron velocity is not constant along a given field line, namely a surface of constant $\Psi$; the second term is the advection of $B_y$ should the electron number density be not constant and the final term is the Ohmic decay term. The equation describing the evolution of $\Psi$ (\ref{dPsi}) has only two terms in the right hand-side:  the first one describing the advection of $\Psi$ by the current on the $x-z$ plane and the second one its Ohmic decay. 

There are two advective terms in the equation, one arising from the density stratification and another due to the impact of the $B_y$ field on $\Psi$. In this work we explore a model of minimal complexity to demonstrate the improvement achieved by the application of an Upwind scheme. Thus, we assume that $n_{\rm e}={\rm const.}$ allowing for the following advective velocity for $\Psi$:
 \begin{align}
    {\mathbf v}_{adv}=\frac{c}{4\pi n_{\rm e} e}\left(\frac{\partial B_y}{\partial z} \hat{\mathbf x} -\frac{\partial B_y}{\partial x} \hat{\mathbf z}\right) 
    \label{ADV}
\end{align}
This advective velocity is dot-producted with the gradient of $\Psi$, providing the first term of equation (\ref{dPsi}). This term implies that any information related to the evolution travels in the direction of the advective term. Thus, if a central difference scheme is used for the evaluation of the first derivatives it would be prone to numerical instabilities, as it could take into account downstream points where the information about the advective field has not arrived yet. On the contrary, the use of a Upwind scheme takes into account only the points where the information has arrived and is anticipated to be more stable numerically.

\section{Numerical setup and strategy}
\label{Num_setup}
The numerical integration of differential equations, even in 1-D systems, containing advective terms is greatly enhanced by the use of Upwind schemes. In what follows we demonstrate the implementation of an Upwind scheme for the numerical solution of equations (\ref{dBy}) and (\ref{dPsi}) and we compare it against the integration with the usage of central difference derivatives. 

{\bf }We set 
\begin{eqnarray}
\tilde{\bf B}=\frac{c}{4 \pi e n_e L^2} {\bf B}\,,
\end{eqnarray}

allowing to the normalisation of equation (\ref{HALL}) as follows:
\begin{equation}
    \frac{\partial{\mathbf{\tilde B}}}{\partial{\tilde t}}= -\nabla \times \left[(\nabla \times \mathbf{\tilde B})\times \mathbf{\tilde B} +R_{H}^{-1} \nabla \times \mathbf{\tilde B}\right]
    \label{HALL_NORM}
\end{equation}
where time $\tilde t$ is measured in Hall-timescale $t_{H}$ units for a magnetic field of unit strength. In what follows we shall use the normalised magnetic field and drop the tilde. 

\subsection{Discretisation}

We consider a square integration domain of unit side with $x\times z\in [0,1]\times[0,1]$, that we have descretised for $i \in \{0, 1, \dots, n_x\}$ and $j\in \{ 0, 1, \dots, n_z\}$. We denote time with an upper index and position with lower indices, with $x=i \delta x$ and $z=j \delta z$, where $\delta x=1/n_x$ and $\delta z=1/n_{z}$.

We solve numerically the equations (\ref{dBy}) and (\ref{dPsi}) using two methods: a first order Forward in Time and Central in Space scheme (FTCS) and an Upwind scheme \citep{1980wdch.book.....P}. 

We define the following operators for a scalar quantity $Q_{i,j}^n$:
\begin{eqnarray}
D_t\left( Q^n_{i,j}\right) \equiv \frac{Q_{i,j}^{n+1}-Q_{i,j}^n}{\delta t}
\end{eqnarray}
\begin{eqnarray}
L \left(Q_{i,j}^n\right) \equiv \frac{Q_{i+1,j}^n+Q_{i-1,j}^n+Q_{i,j+1}^n+Q_{i,j-1}^n-4Q_{i,j}^n}{\delta x \delta z}
\end{eqnarray}
\begin{eqnarray}
D_x \left(Q_{i,j}^n\right) \equiv \frac{Q_{i+1,j}^n-Q_{i-1,j}^n}{2\delta x}
\end{eqnarray}
\begin{eqnarray}
D_z \left(Q_{i,j}^n\right) \equiv \frac{Q_{i,j+1}^n-Q_{i,j-1}^n}{2\delta z}\,.
\end{eqnarray}
Using these operators, equations (\ref{dBy}), once normalised and discretised, takes the following form:
\begin{eqnarray}
D_{t} \left(B_{y~i,j}^{n}\right) = D_{z} (L (\Psi_{i,j}^{n})) D_{x}(\Psi_{i,j}^{n}) \\
-D_{x} (L (\Psi_{i,j}^{n})) D_{z}(\Psi_{i,j}^{n})\nonumber \\
+R^{-1}_{H} L(B_{y~i,j}^{n}).
\end{eqnarray}
This form of the equation is applicable on both FTCS and Upwind schemes. On the contrary, equation (\ref{dPsi}) contains an advective term and its form is different in the FTCS and Upwind scheme. In the FTCS scheme the equation takes the form: 
\begin{eqnarray}
D_{t}(\Psi_{i,j}^{n})=-D_{z}(B_{y}{}_{i,j}^{n})D_{x}(\Psi_{i,j}^{n})+D_{x}(B_{y}{}_{i,j}^{n})D_{z}(\Psi_{i,j}^{n}) \\
+R_{H}^{-1}L(\Psi_{i,j}^{n}).
\end{eqnarray}
The expression differs, however, when the Upwind scheme is applied. In particular, from the definition of equation (\ref{V_ADV}) if the term $D_{z}(B_{y}{}_{i,j}^{n})>0$ we obtain $v_{adv,x}<0$ and vice versa. Similarly, if $D_{x}(B_{y}{}_{i,j}^{n})>0$ then $v_{adv,z}>0$, notice the different signs of the $y$ and $z$ derivatives of the respective terms. Because of this, we need further to define four more operators: 
\begin{eqnarray}
D_{x-}Q_{i,j}^n\equiv  \frac{Q^{n}_{i,j}-Q^{n}_{i-1,j}}{\delta x}
\end{eqnarray}
\begin{eqnarray}
D_{x+}Q_{i,j}^n\equiv  \frac{Q^{n}_{i+1,j}-Q^{n}_{i,j}}{\delta x}
\end{eqnarray}
\begin{eqnarray}
D_{z-}Q_{i,j}^n\equiv  \frac{Q^{n}_{i,j}-Q^{n}_{i,j-1}}{\delta z}
\end{eqnarray}
\begin{eqnarray}
D_{z+}Q_{i,j}^n\equiv  \frac{Q^{n}_{i,j+1}-Q^{n}_{i,j}}{\delta z}\,.
\end{eqnarray}
Under these considerations the form of equation (\ref{dPsi}) takes one of the following forms depending on the sign of the advective velocity. For $v_{adv,x}>0$ and $v_{adv,z}>0$
\begin{eqnarray}
D_{t} (\Psi^{n}_{i,j})=-D_{z}(B_{y}{}_{i,j}^{n})D_{x+}(\Psi_{i,j}^{n})+D_{x}(B_{y}{}_{i,j}^{n})D_{z+}(\Psi_{i,j}^{n})\nonumber \\
+R^{-1}_H L(\Psi_{i,j}^n)\,.
\end{eqnarray}
For $v_{adv,x}>0$ and $v_{adv,z}<0$
\begin{eqnarray}
D_{t} (\Psi^{n}_{i,j})=-D_{z}(B_{y}{}_{i,j}^{n})D_{x+}(\Psi_{i,j}^n)+D_{x}(B_{y}{}_{i,j}^{n})D_{z-}(\Psi_{i,j}^n)\nonumber \\
+R^{-1}_H L(\Psi_{i,j}^n)\,.
\end{eqnarray}
For $v_{adv,x}<0$ and $v_{adv,z}>0$
\begin{eqnarray}
D_t (\Psi^{n}_{i,j})=-D_{z}(B_{y}{}_{i,j}^{n})D_{x-}(\Psi_{i,j}^n)+D_{x}(B_{y}{}_{i,j}^{n})D_{z+}(\Psi_{i,j}^n)\nonumber \\
+R^{-1}_H L(\Psi_{i,j}^n)\,.
\end{eqnarray}
And finally, for $v_{adv,x}<0$ and $v_{adv,z}<0$
\begin{eqnarray}
D_t (\Psi^{n}_{i,j})=-D_{z}(B_{y}{}_{i,j}^{n})D_{x-}(\Psi_{i,j}^n)+D_{x}(B_{y}{}_{i,j}^{n})D_{z-}(\Psi_{i,j}^n)\nonumber \\
+R^{-1}_H L(\Psi_{i,j}^n)\,.
\end{eqnarray}

We have further applied periodic boundary conditions. As our scheme needs two grid points on either side to evaluate the third derivatives, we have used two sets of ghost points on all four sides of the grid. There, we copied the values at the end of the integration loops as follows:
\begin{eqnarray}
Q^n_{-2,j}&=&Q^n_{n_x-1,j}\nonumber \\
Q^n_{-1,j}&=&Q^n_{n_x,j}\nonumber \\
Q^n_{n_x+1,j}&=&Q^n_{0,j}\nonumber \\
Q^n_{n_x+2,j}&=&Q^n_{1,j}\nonumber \\
Q^n_{i,-2}&=&Q^n_{i,n_z-1}\nonumber \\
Q^n_{i,-1}&=&Q^n_{i,n_z}\nonumber \\
Q^n_{i,n_z+1}&=&Q^n_{i,0}\nonumber \\
Q^n_{i,n_z+2}&=&Q^n_{i,1} 
\end{eqnarray}

In our simulations we have used a variable time step, set by the maximum electron velocity in the system. In systems with weak magnetic fields, we switch-off this condition as this would give rapid dissipation and lead to numerical divergence.

\subsection{Initial Conditions}

We explore the impact of the Upwind method on convergence subject to two main parameters: the ratio of planar to vertical field and the value of the Hall parameter $R_{H}$. The structure of the magnetic field is the same, given by the following expressions:
\begin{eqnarray}
B_{y}&=&B_{0} x^2z^2(x-1)(z-1)\,,  \nonumber\\
\Psi&=&\Psi_{0} xz(x-1)(z-1)\,.
\end{eqnarray}
The expression for $B_y$ has a peak at $(2/3,2/3)$ whereas the one for $\Psi$ peaks at $(1/2,1/2)$.  We note that the initial conditions for $B_y$ and $\Psi$ have the same values for $x=0$ and $x=1$; and $z=0$ and $z=1$, and the derivatives:
\begin{eqnarray}
\frac{\partial B_y}{\partial x}|_{z=0}=\frac{\partial B_y}{\partial x}|_{z=1}=0\,,\nonumber \\
\frac{\partial B_y}{\partial z}|_{x=0}=\frac{\partial B_y}{\partial z}|_{x=1}=0\,,
\end{eqnarray} 
implying that the electron fluid velocity normal to the boundaries of the domain is initially equal to $0$.  

In the limit where $B_y$ dominates the evolution and high conductivity, the system will evolve towards a state where the contours of constant $\Psi$ will tend to coincide with the ones with constant $B_{y}$. If conductivity becomes weaker, $B_y$ will dissipate in the system and the magnetic field energy will decrease, while the components of the field will try to obtain an advective-diffusive equilibrium similar to the Hall attractor \citep{2014PhRvL.112q1101G}. For systems where the $B_y$ component is comparable to the planar components the system will have a more complex evolution due to the interplay of the currents due to $\Psi$ and their impact on $B_y$. 

In our simulations we have set the value of $B_0=105$ so that the integral of the magnetic field energy equals:
\begin{eqnarray}
E_y=\frac{1}{8 \pi}\int B_y^2 dx dz=\frac{1}{8\pi}\,.
\end{eqnarray}
We further define the energy in the planar components $B_x$ and $B_{z}$ in a similar manner:
\begin{eqnarray}
E_{xz}=\frac{1}{8 \pi}\int \left(B_x^2+B_z^2\right) dx dz\,.
\end{eqnarray}
We vary the value of $\Psi_0$ so that the ratio of the energy corresponding to the $B_x$ and $B_z$ component over the $B_y$ energy to be $0.01$, $0.1$, $0.5$ and $1$, by setting $\Psi_0$ equal to $0.653$, $2.067$, $4.62$ and $6.53$ . Obviously, a state where the planar field is identically zero $\Psi=0$ leads to a system that does not evolve other than for the decay of $B_y$.

\subsection{Convergence Criteria}

To successfully compare the two schemes we use the following convergence criteria. The magnetic field energy decays solely due to Ohmic effect \citep{2002MNRAS.337..216H}, with the following equation describing the magnetic energy decay:
\begin{equation}
    \frac{\partial}{\partial t}\left(\frac{1}{2}\int_{V}\mathbf{B}^2dV\right)=-\frac{1}{R_H}\int_{V}\mathbf{j}^2dV\,.
    \label{dEner}
\end{equation}
Thus, our fist convergence criterion is that the energy in the magnetic should monotonically decrease. 
Moreover, the same electric current distribution  leads to a slower magnetic field decay for runs if a higher $R_H$ is chosen. Nevertheless, given the non-linearity of the evolution, the electric currents later on, will not be identical for runs with different values of $R_H$, thus a direct comparison may not be possible.\par We studied the difference given by the terms
\begin{equation}
m=\frac{|\frac{\partial}{\partial t}\left(\frac{1}{2}\int_{V}\mathbf{B}^2dV\right)+\frac{1}{R_H}\int_{V}\mathbf{j}^2dV|}{\frac{1}{R_H}\int_{V}\mathbf{j}^2dV} \end{equation} 
for several runs (C001 200, U001 200, C010 100, U010 100, C050 100, U050 100),  and we find that it remains within a few percent. We note however that there is a rising trend of this value with higher $R_H$, indicative of the impact of the numerical dissipation for higher Hall parameters that we have already noted. 

 We note that a further criterion that can be used in this context \cite{Wareing:2010} is the evolution of the magnetic helicity which is given by the following expression:
\begin{eqnarray}
\frac{\partial}{\partial t} \left(\int_V {\bf A}\cdot {\bf B} ~dV\right)=-\frac{1}{R_H}\int_V {\bf B}\cdot {\bf j}~dV\,,
\end{eqnarray}
where ${\bf A}$ is the vector potential so that $\nabla \times {\bf A}={\bf B}$
Given that our code integrates directly the quantities $B_y$ and $\Psi$ the evaluation of ${\bf A}$ will require a further integration, as its $A_{x}$ and $A_{z}$ components are related to integrals of $B_y$ and this would introduce further numerical errors. So we confine ourselves to our first criterion related to the energy which is more accurate.

A second criterion that we impose for convergence is the size of the time-step. In the simulations, we use a Courant condition \citep{https://doi.org/10.1002/cpa.3160050303} by setting the size of the timestep $\delta t=0.1~ C_P~ \delta x~ \delta z$, where $C_P=\frac{1}{{\rm max}\{ |{\bf v}_e|, 0.5\}}$. The electron velocity is evaluated from equation (\ref{V_ADV}) with a central difference scheme for the derivatives of $\Psi$ and $B_y$. This choice slows down the integration, once strong currents form, and thus electron velocities,  to prevent numerical divergence. However, if a numerical instability occurs leading to high values of the derivatives, it will halt the evolution of the system by enforcing  a very small $\delta t$. We have also used a base value for the electron velocity $0.5$, as for very weak magnetic fields, the value of $|{\bf v}_{e}|$ become very small and it would lead to a very large timestep making the Ohmic term numerically unstable. Thus, our second criterion is that the timestep does not become zero.

\label{isosetup}

\section{Simulation Results}
\label{Sim}
\begin{figure*}
a\includegraphics[width=0.235\textwidth]{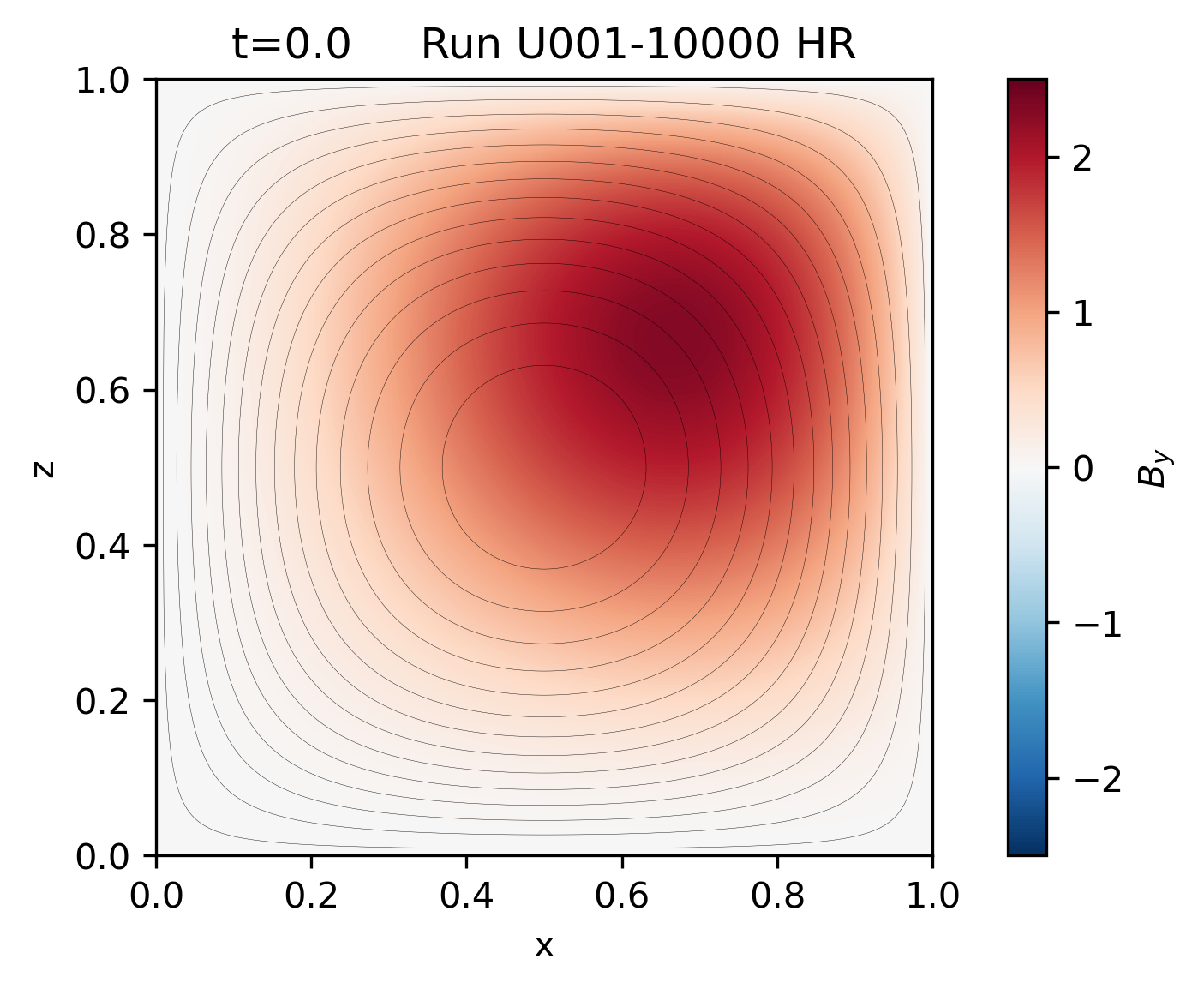}
b\includegraphics[width=0.235\textwidth]{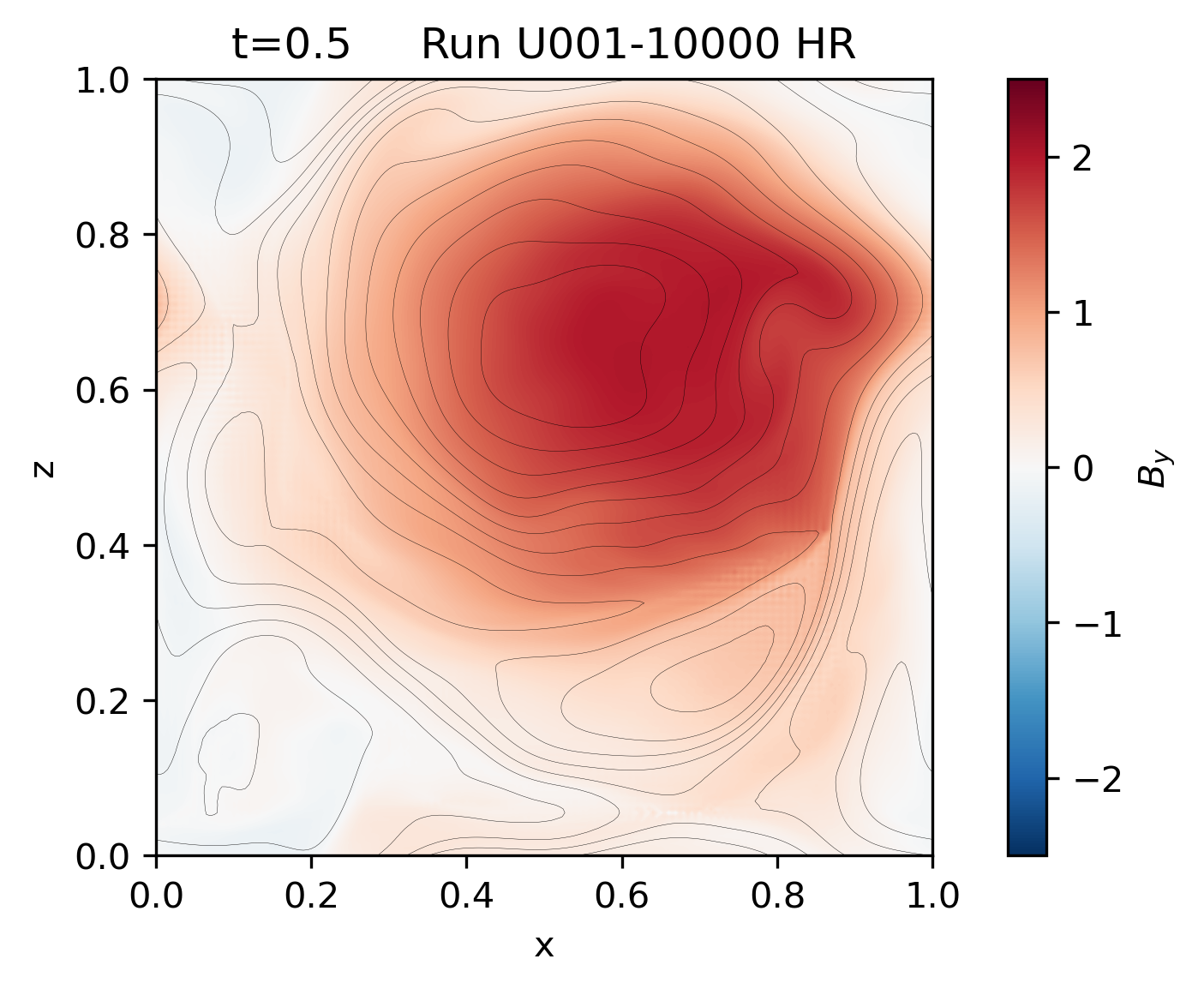}
c\includegraphics[width=0.235\textwidth]{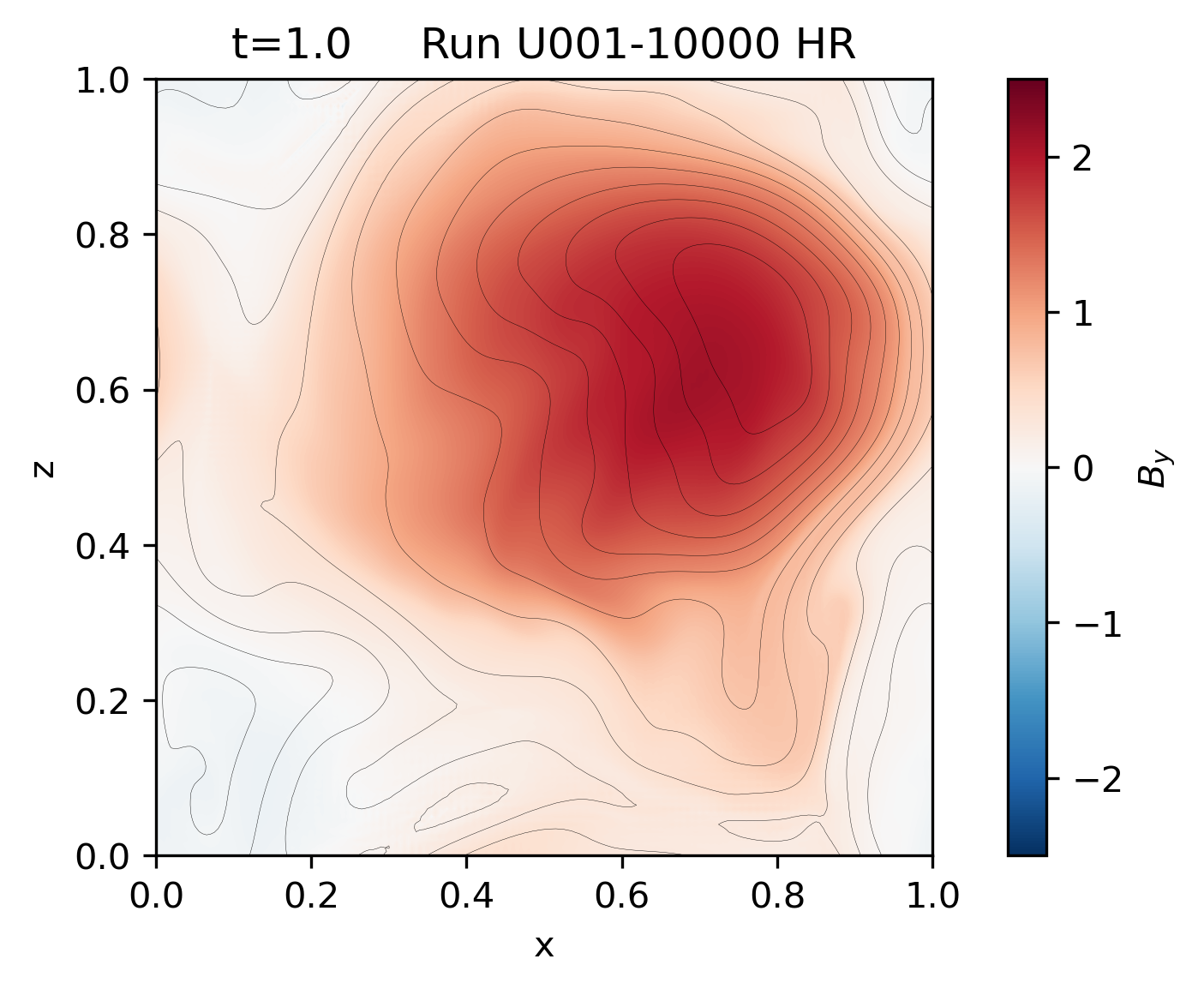}
d\includegraphics[width=0.235\textwidth]{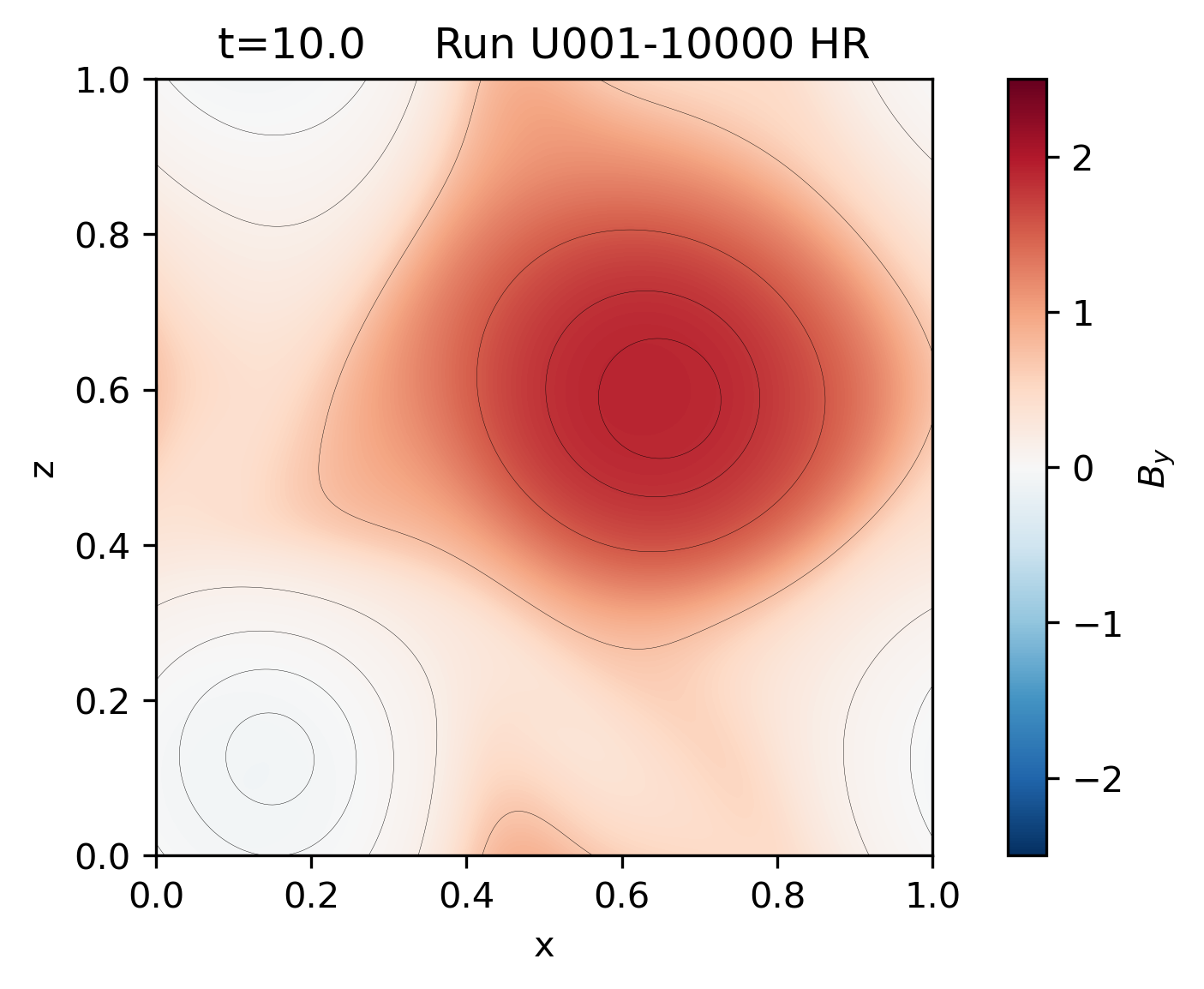}
\caption{Magnetic field structure of the run U001-10000-HR at times $t=0,~0.5~, 1~10$. }
\label{Fig:1}
\end{figure*}

\begin{figure*}
a\includegraphics[width=0.235\textwidth]{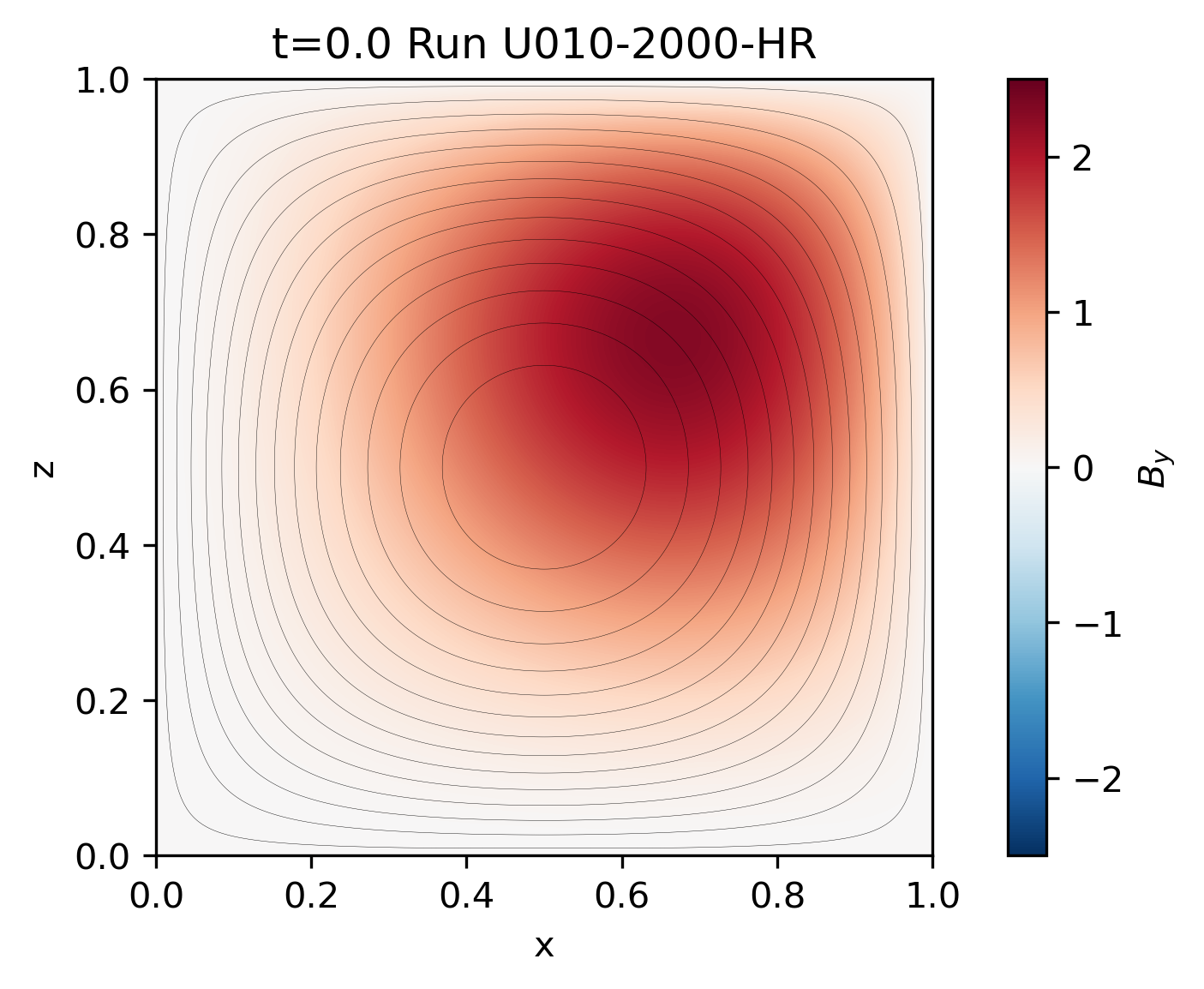}
b\includegraphics[width=0.235\textwidth]{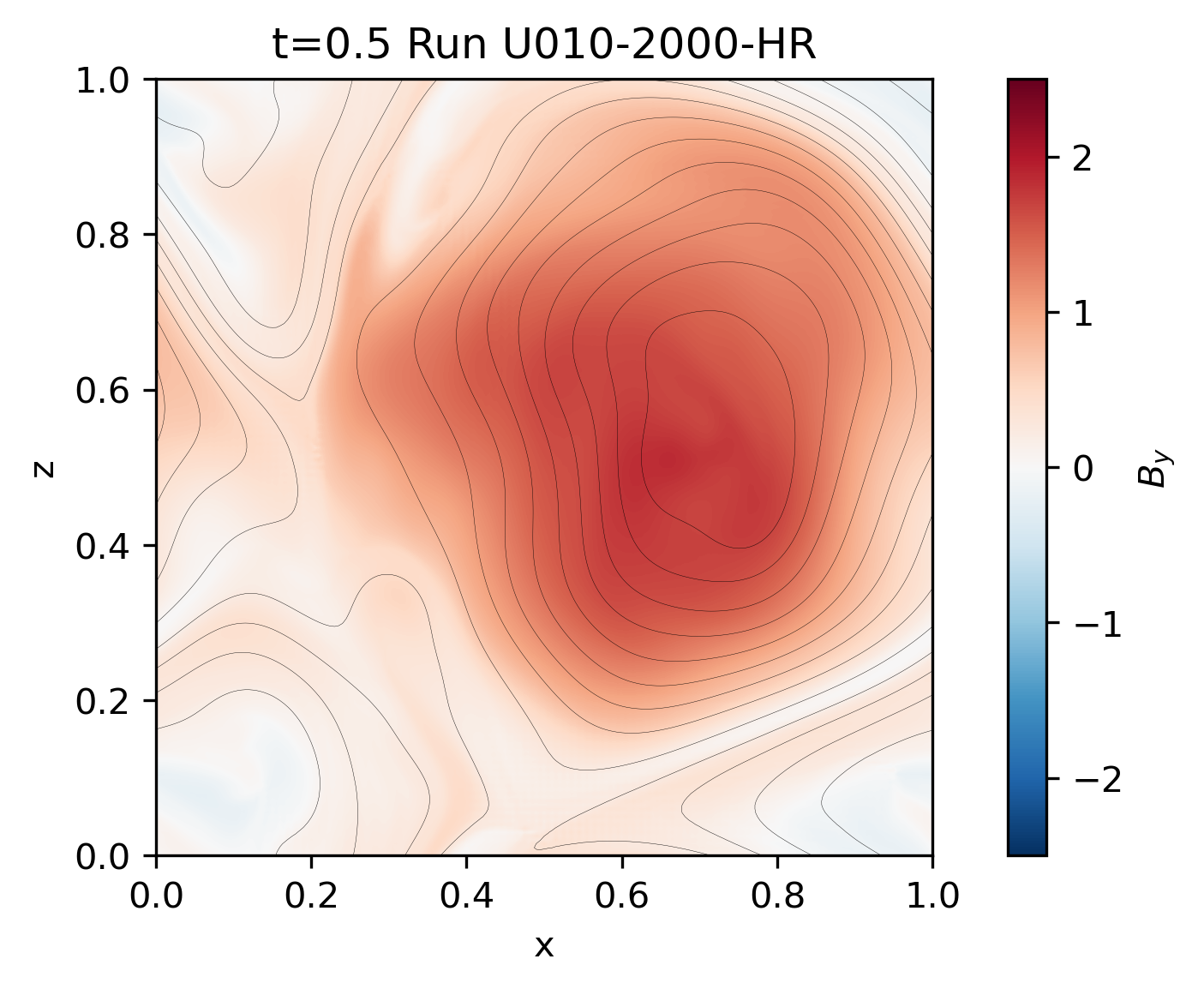}
c\includegraphics[width=0.235\textwidth]{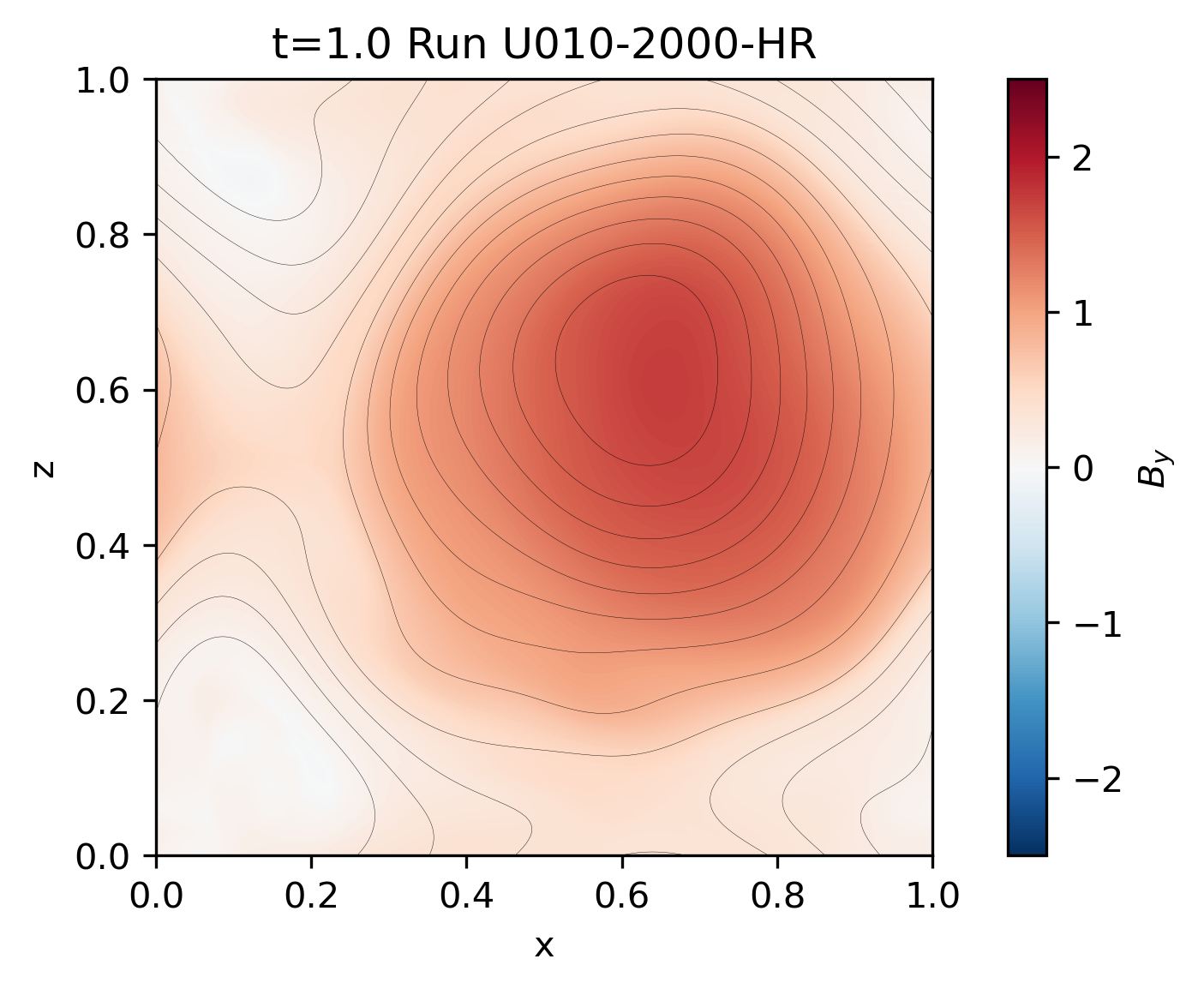}
d\includegraphics[width=0.235\textwidth]{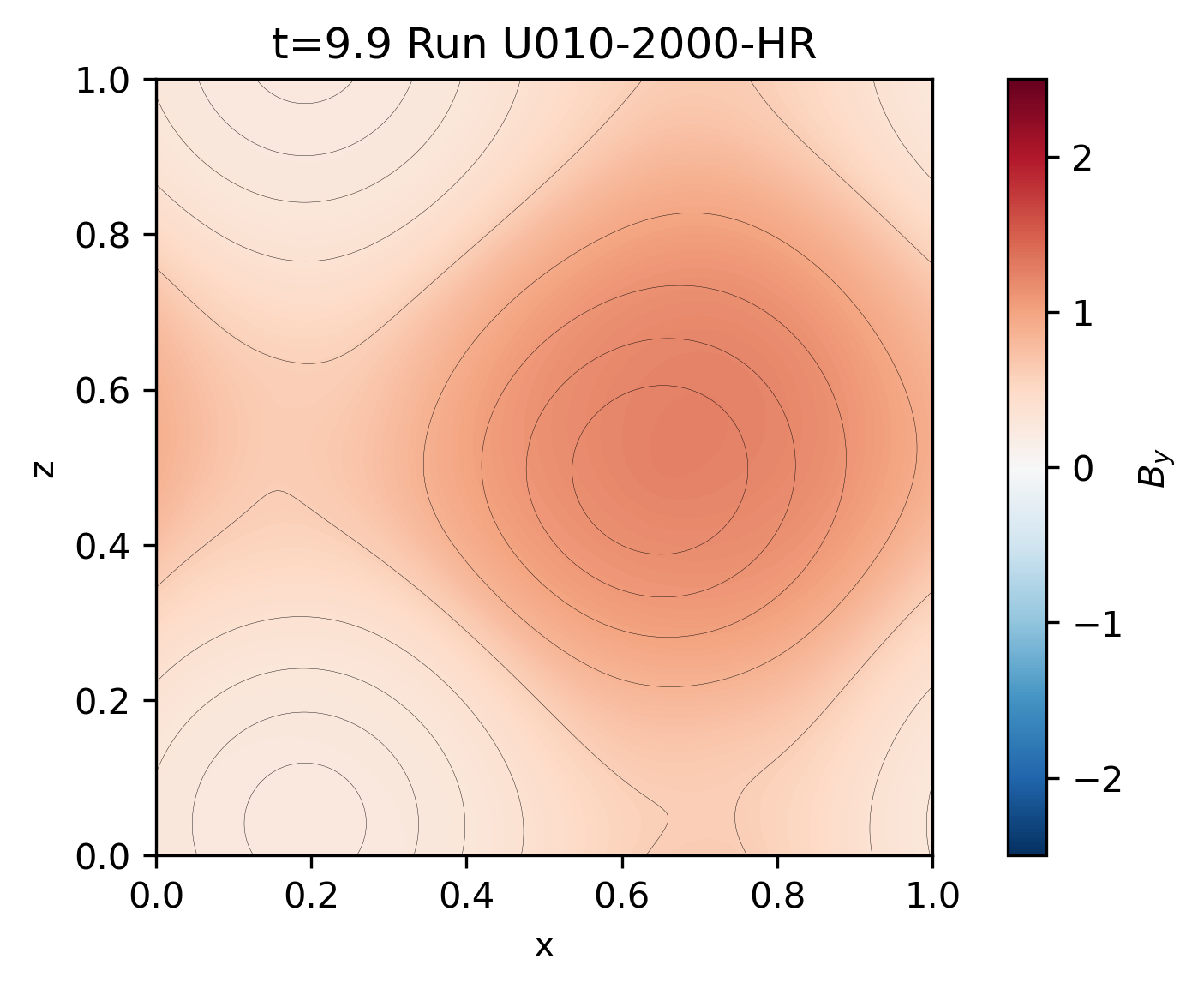}
\caption{Magnetic field structure of the run U010-2000-HR at times $t=0,~0.5~, 1~10$. }
\label{Fig:2}
\end{figure*}

\begin{figure*}
a\includegraphics[width=0.235\textwidth]{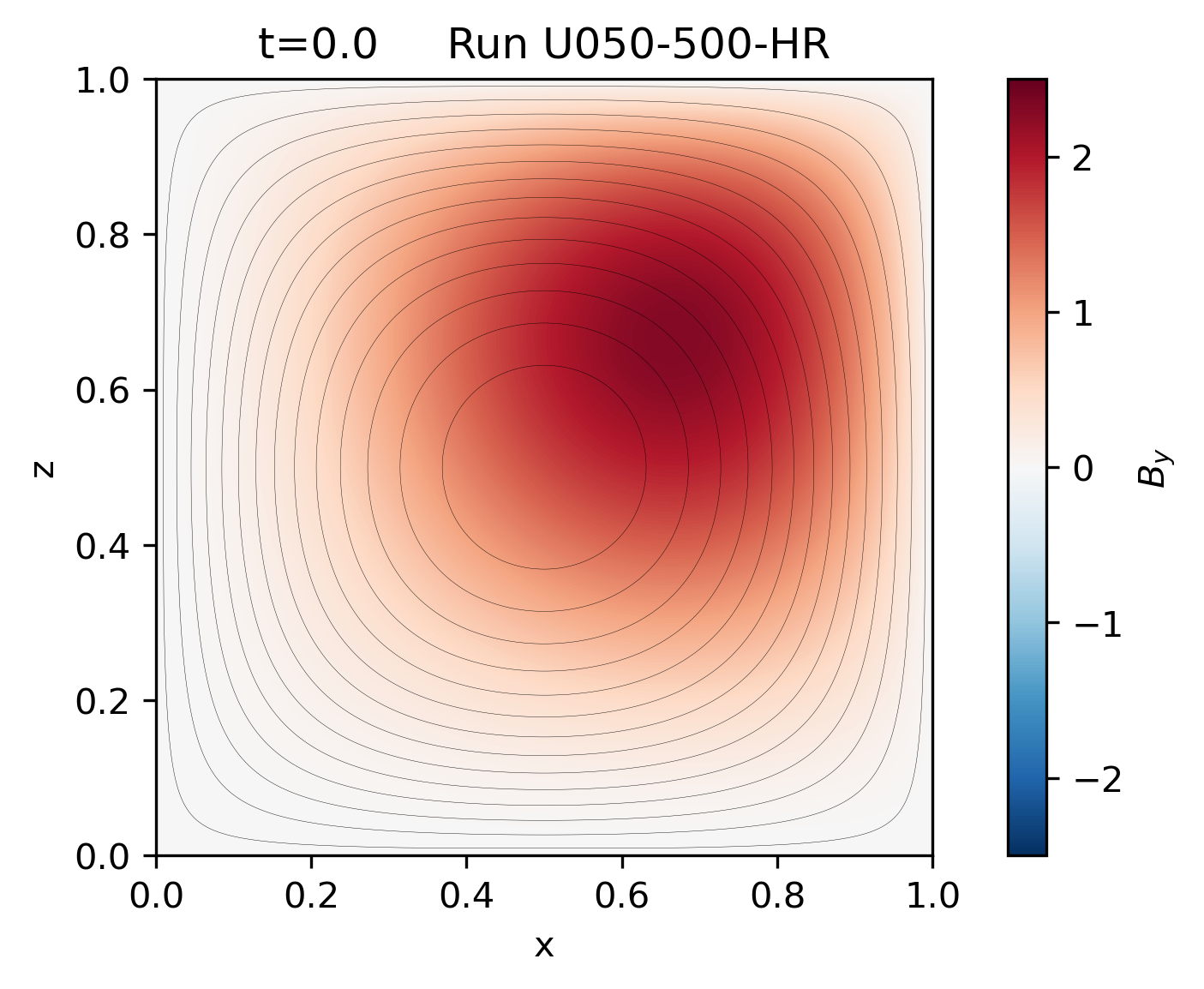}
b\includegraphics[width=0.235\textwidth]{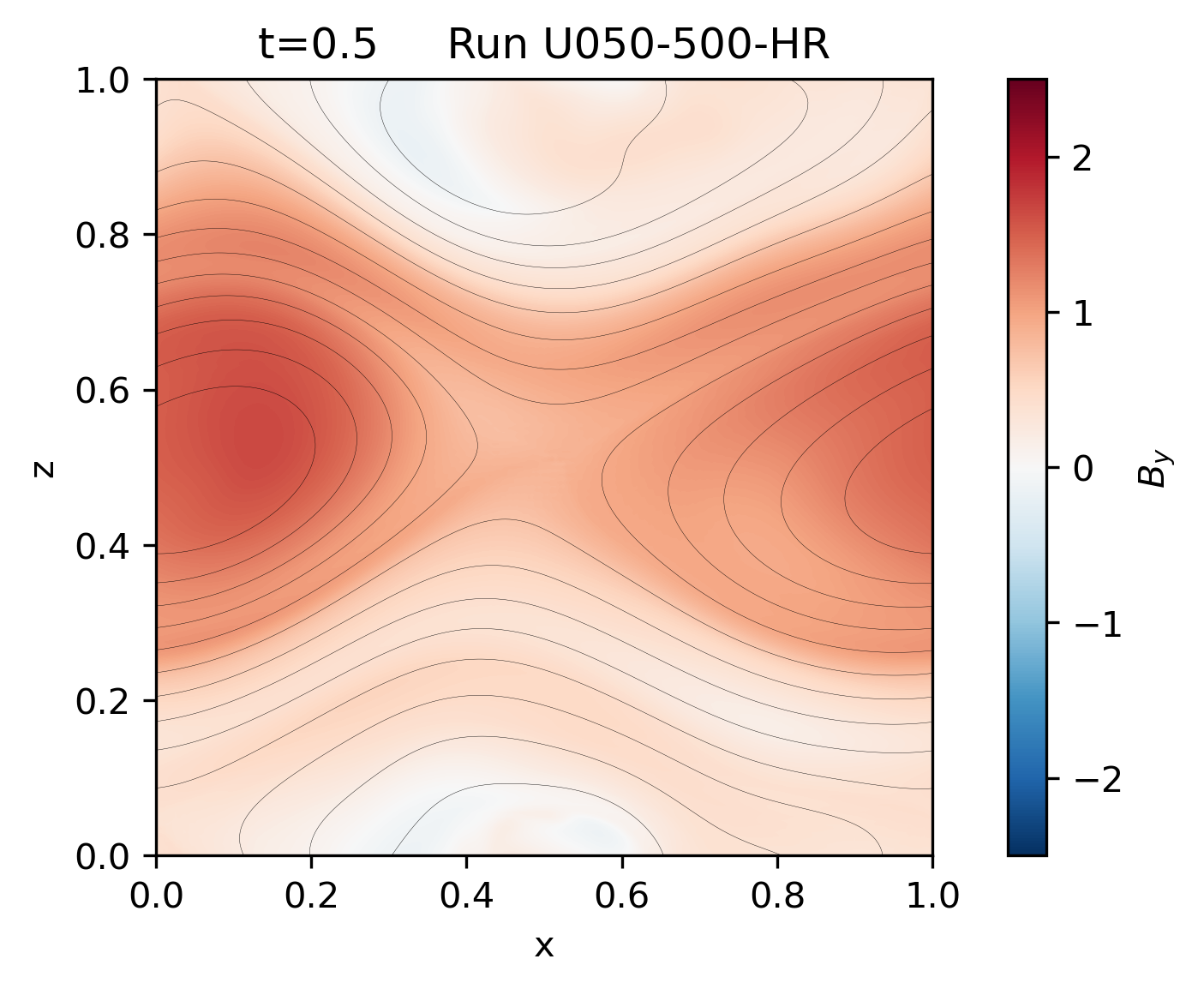}
c\includegraphics[width=0.235\textwidth]{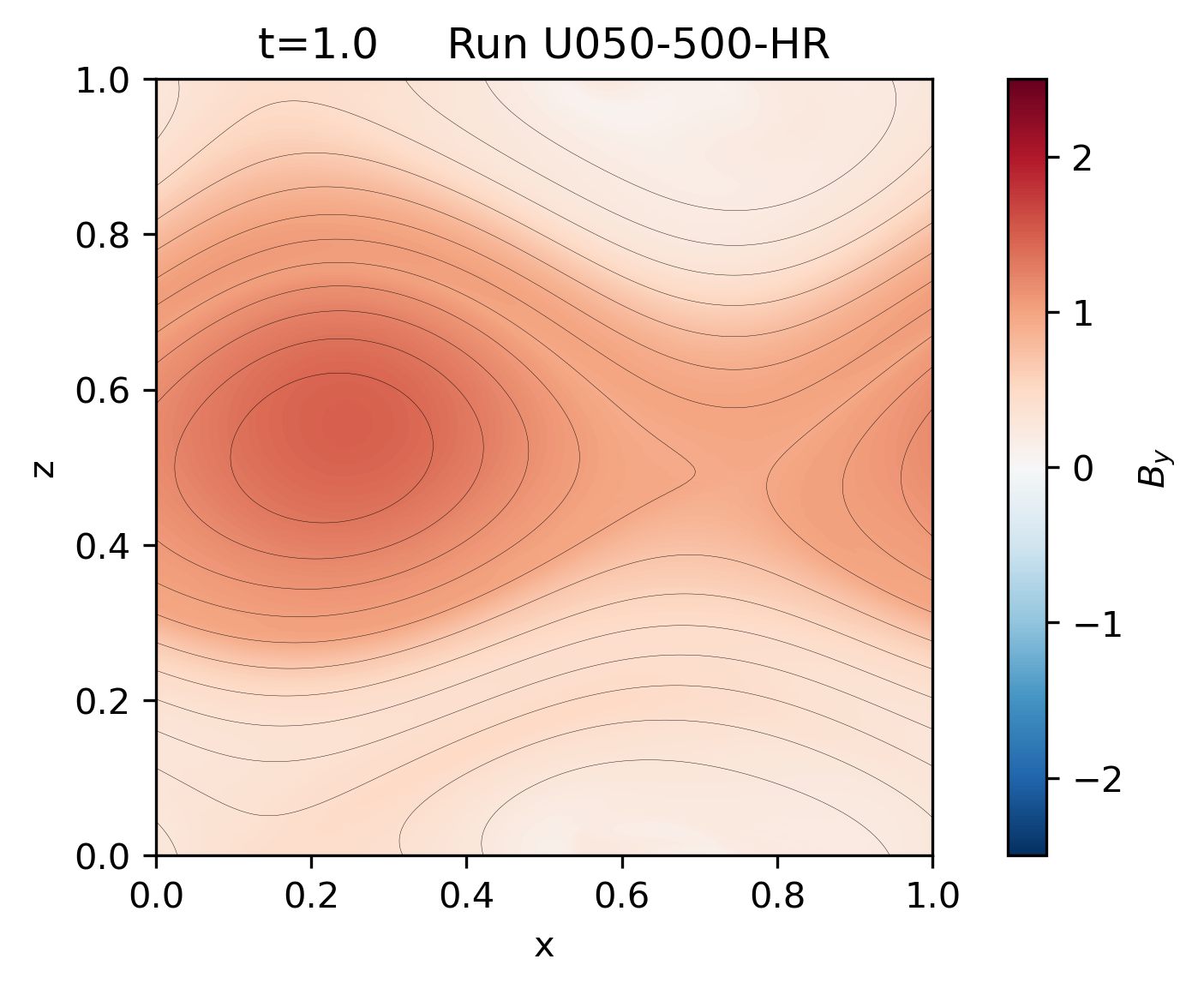}
d\includegraphics[width=0.235\textwidth]{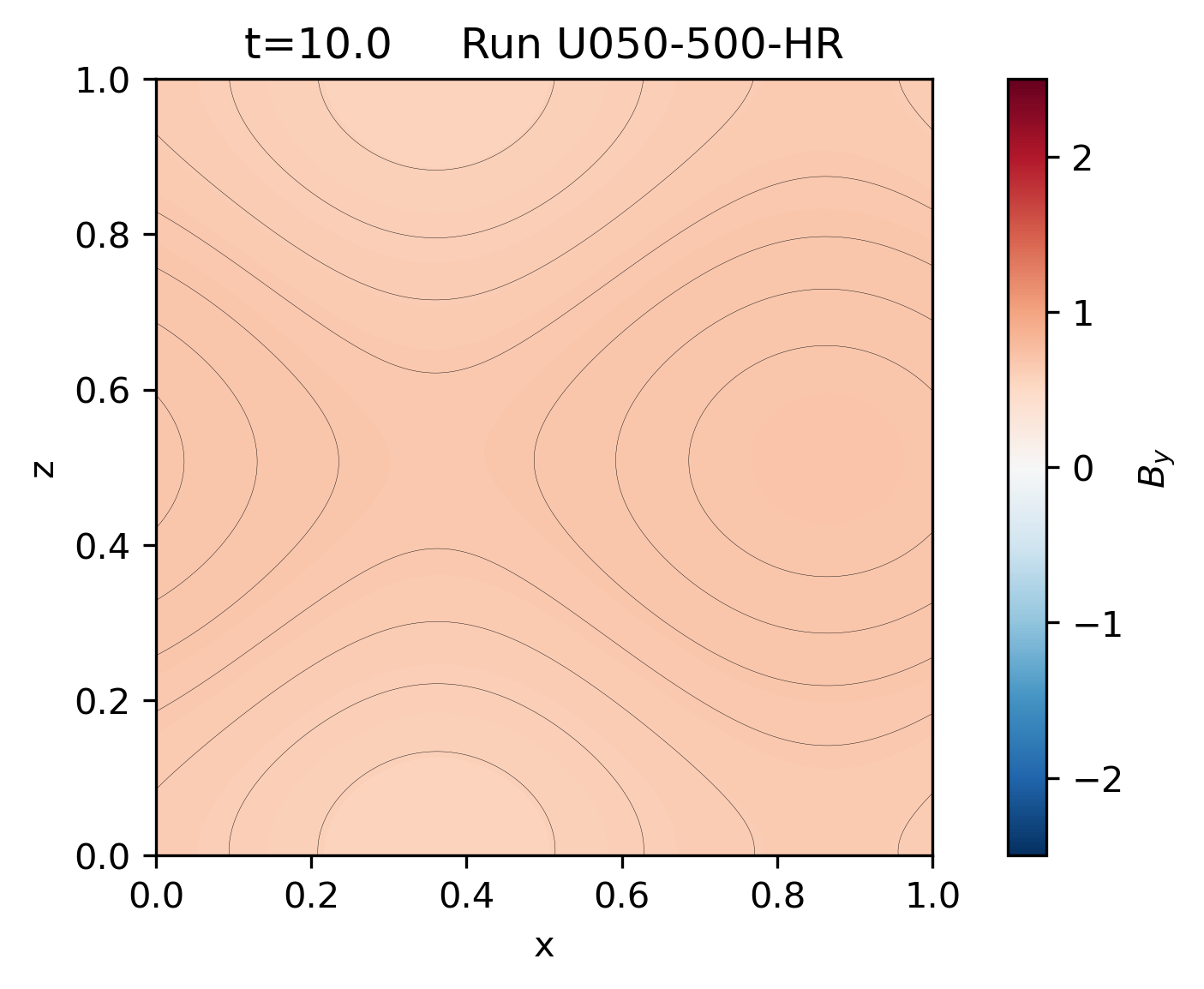}
\caption{Magnetic field structure of the run U050-500-HR at times $t=0,~0.5~, 1~10$. }
\label{Fig:3}
\end{figure*}

\begin{figure*}
a\includegraphics[width=0.235\textwidth]{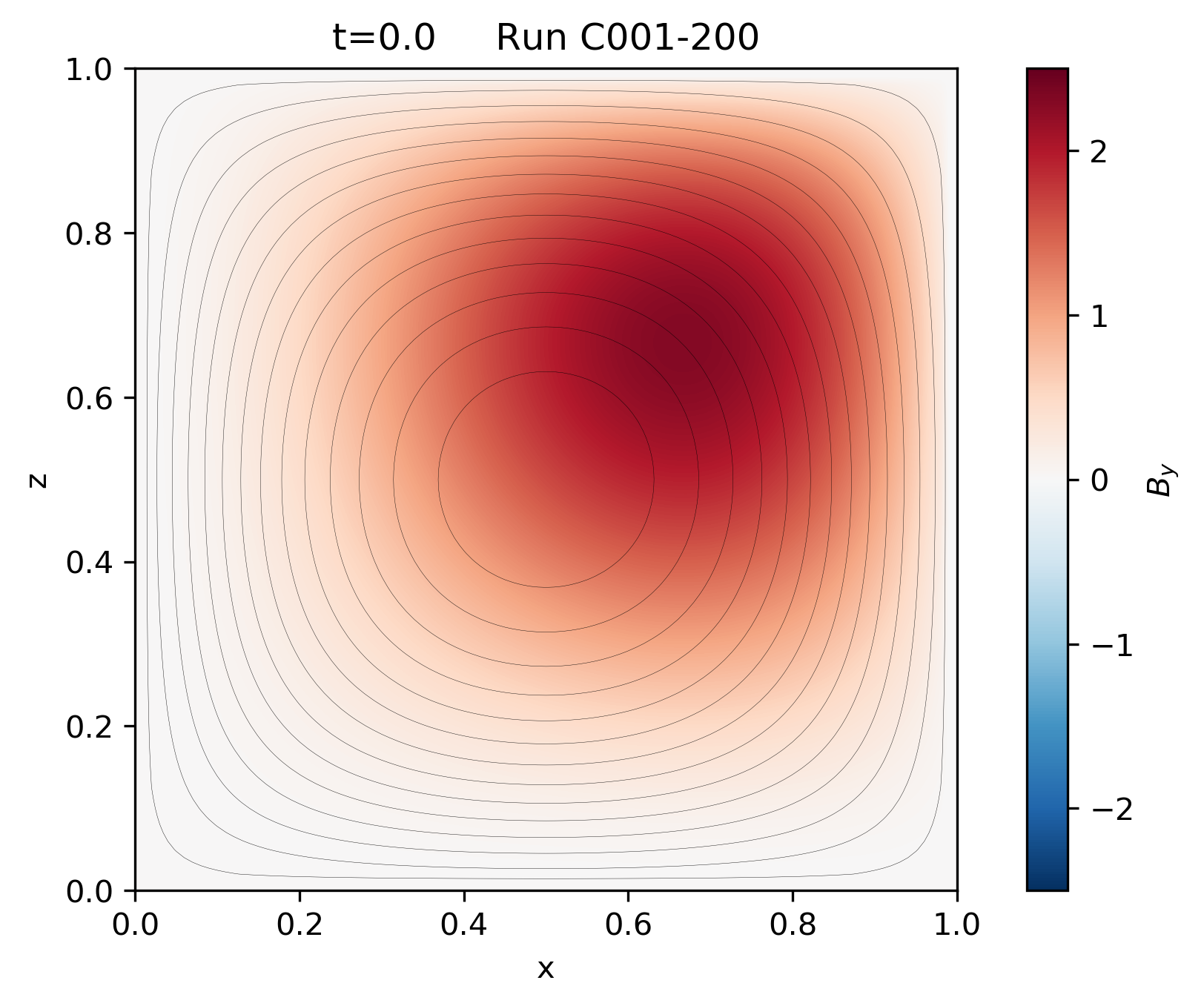}
b\includegraphics[width=0.235\textwidth]{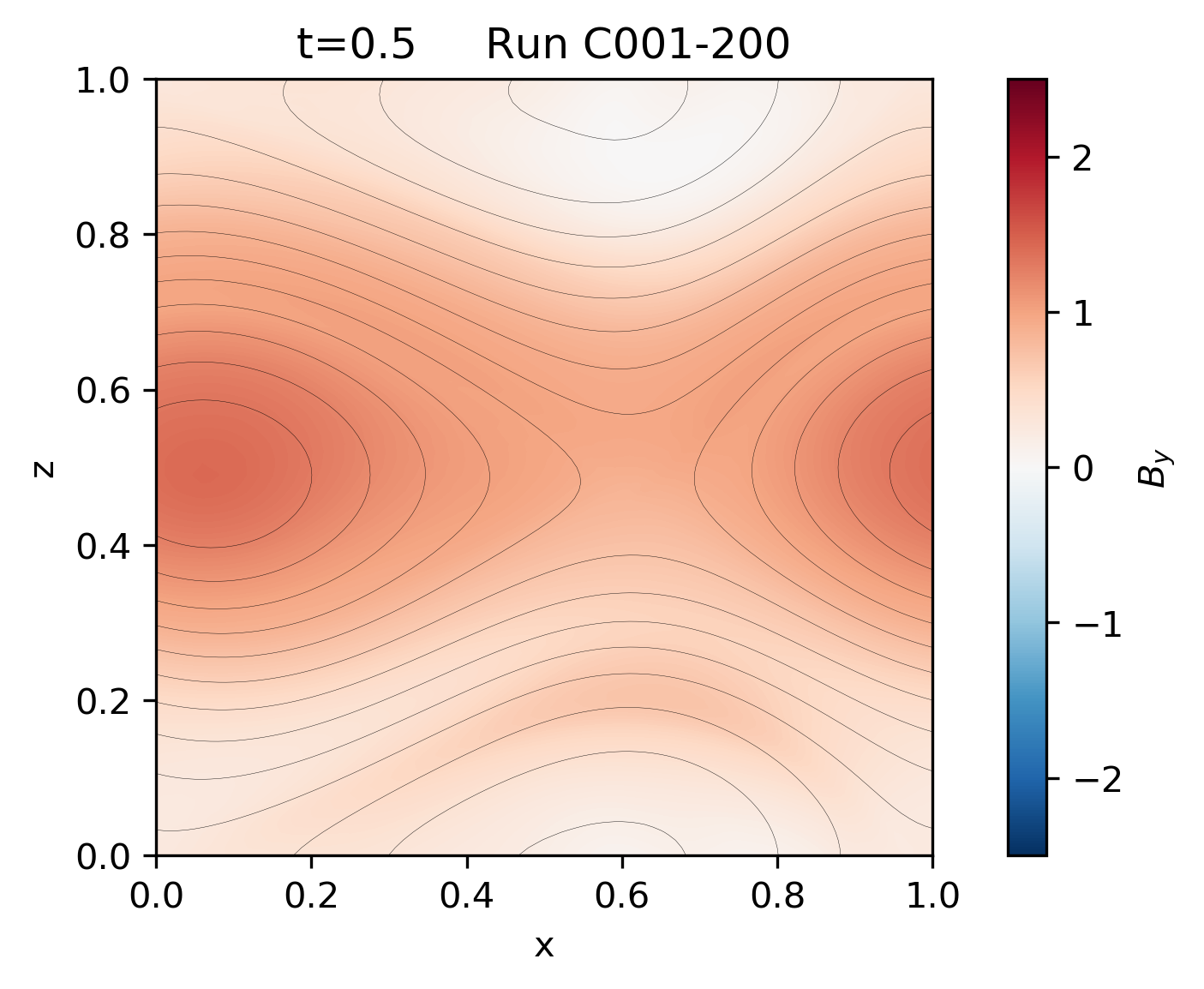}
c\includegraphics[width=0.235\textwidth]{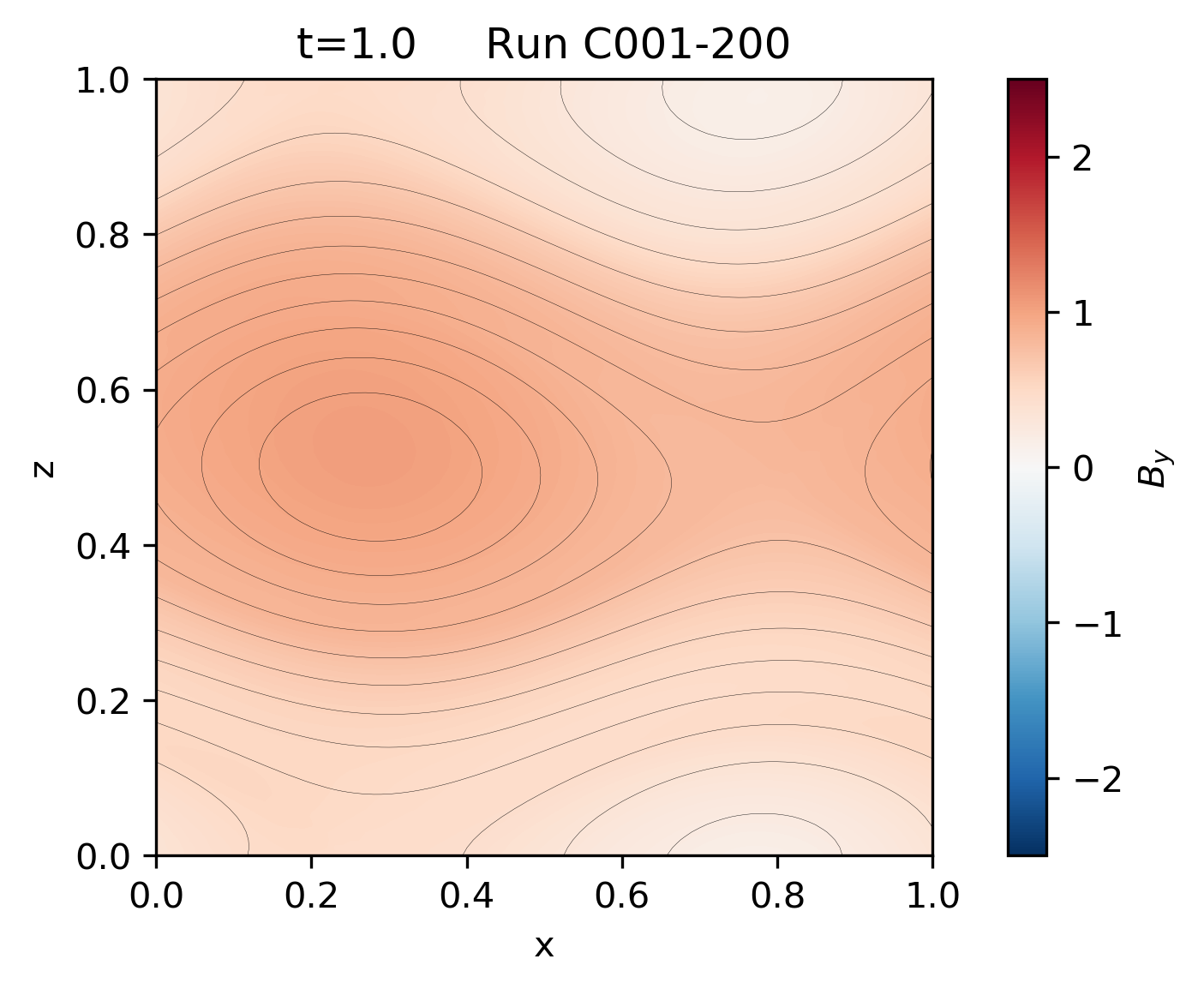}
d\includegraphics[width=0.235\textwidth]{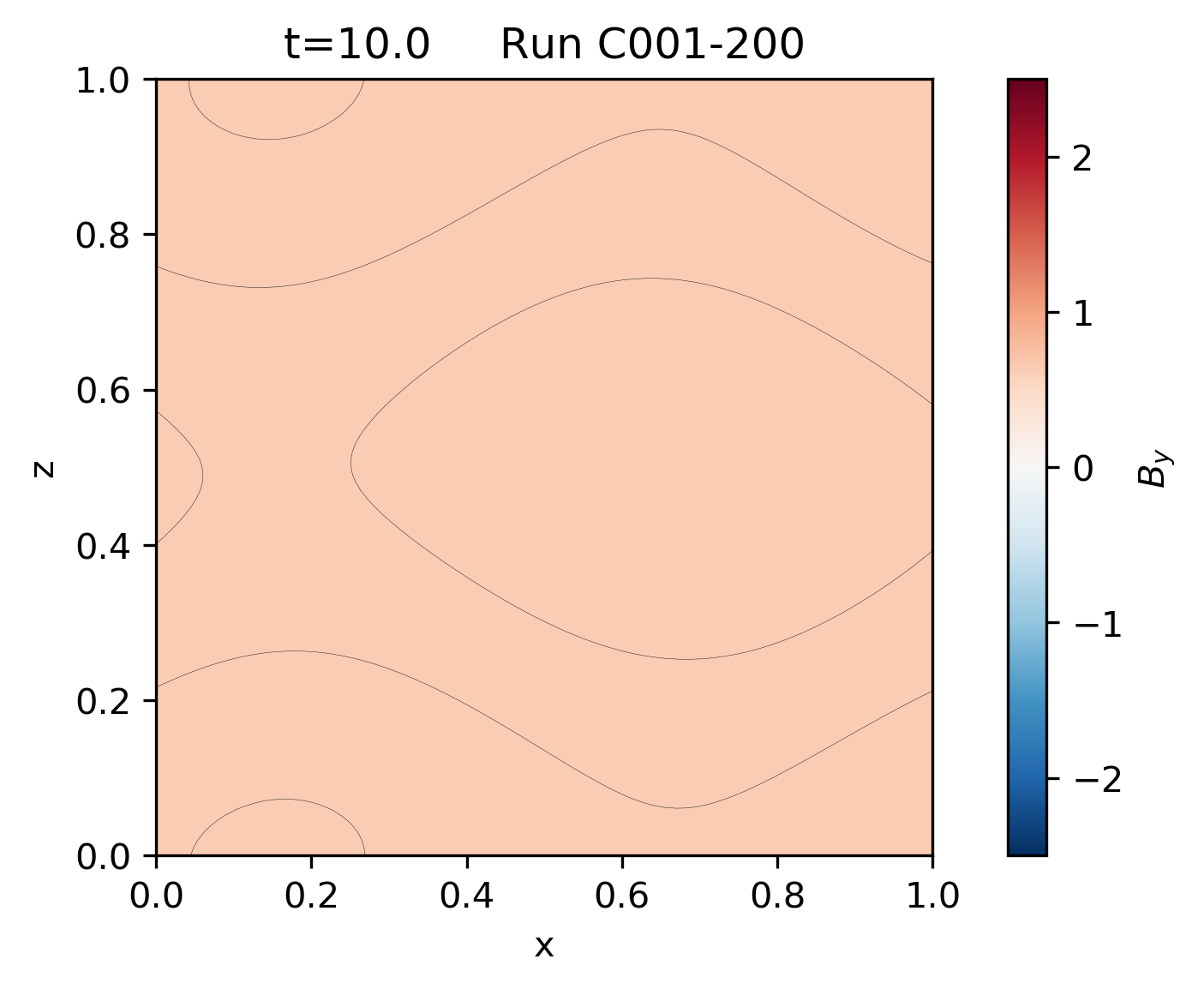}
\caption{Magnetic field structure of the run U100-200 at times $t=0,~0.5~, 1~10$. }
\label{Fig:4}
\end{figure*}

\begin{figure}
a\includegraphics[width=0.48\textwidth]{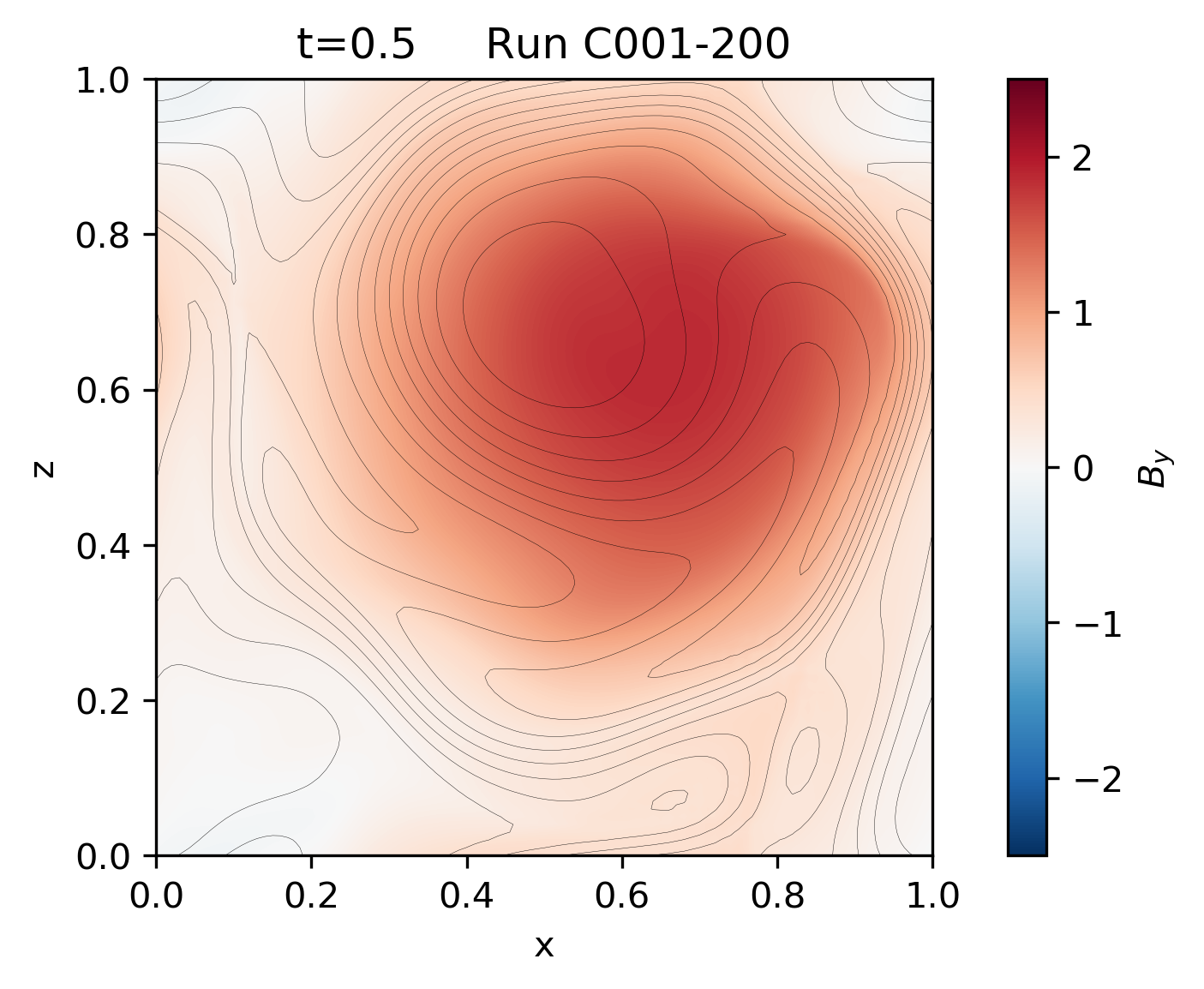}
b\includegraphics[width=0.48\textwidth]{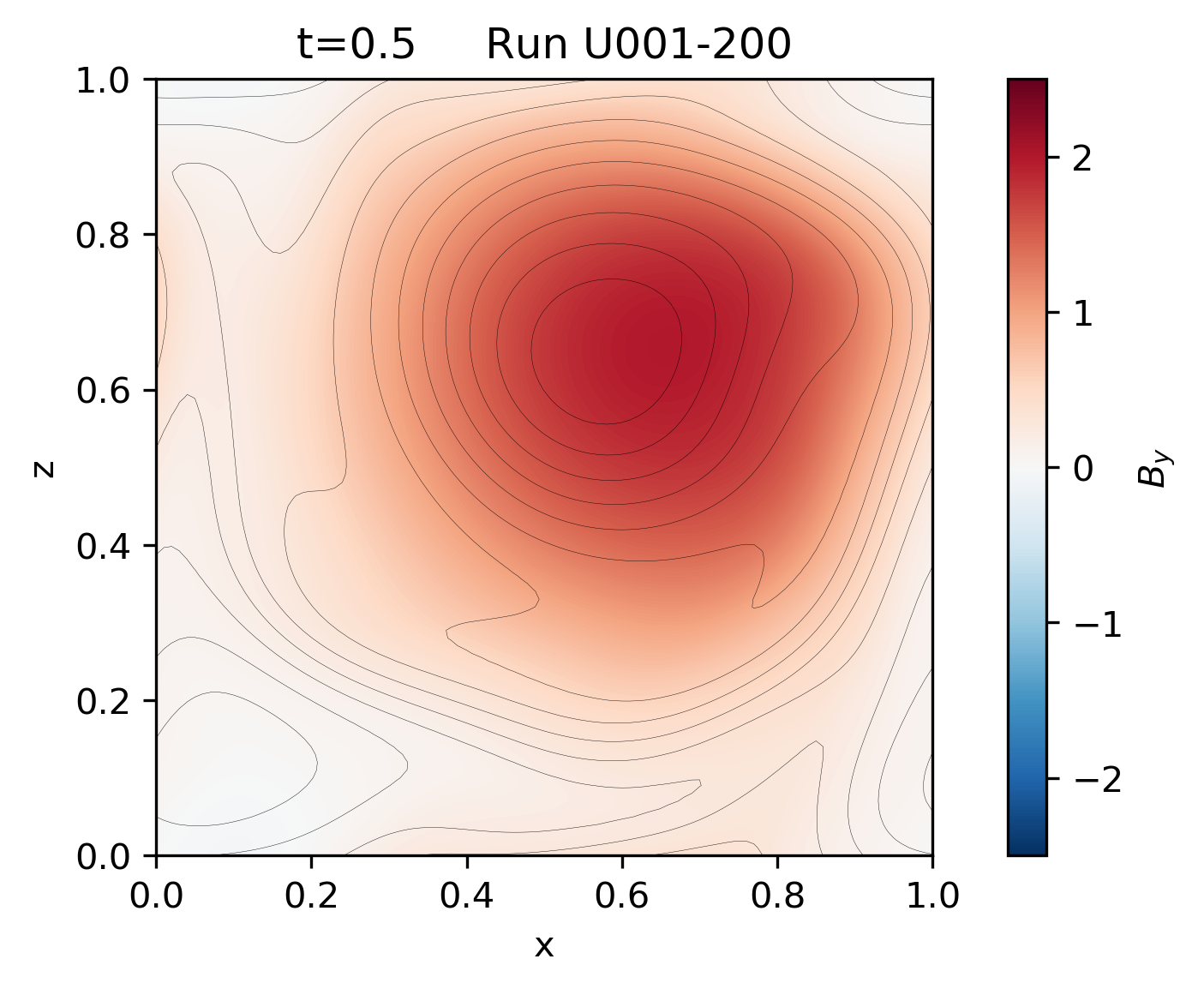}
\caption{Magnetic field structure of the run C001-200 (a) and U001-200 (b) at $t=0.5$}
\label{Fig:5}
\end{figure}

We run identical simulations with initial conditions and physical parameters using both the FTCS and Upwind scheme. In our simulations we have increased progressively the value of $R_H$ until the simulation diverges. To deem a simulation as divergent we require at least one of the two criteria stated in section \ref{isosetup} to be fulfilled. We use as our base resolution $100^2$, we have also run a few simulations with higher resolution $200^2$ to ensure the numerical convergence of the runs and investigate the role of numerical resistivity. We present the parameters used in the simulations and whether the run converge or not in Tables \ref{TAB1}-\ref{TAB4} and stills from the simulations are shown in Figures \ref{Fig:1} -\ref{Fig:5}.  
\begin{table}[h!]
\centering
\begin{tabular}{|l |c |c |c |c |c|} 
 \hline
 Name & $E_{y}$ & $E_{xz}$ & $R_{H}$ & Conv. & Res.\\ [0.5ex] 
 \hline
 C001-50 & 1 & 0.01 & 50 & Yes & $100^2$ \\ 
 C001-100 & 1 & 0.01 & 100 & Yes & $100^2$ \\
 C001-200 & 1 & 0.01 & 200 & Yes & $100^2$ \\
 C001-500 & 1 & 0.01 & 500 & No & $100^2$ \\
 [1ex] 
 \hline
  U001-200 & 1 & 0.01 & 200 & Yes & $100^2$ \\ 
U001-500 & 1 & 0.01 & 500 & Yes & $100^2$ \\
U001-1000 & 1 & 0.01 & 1000 & Yes & $100^2$ \\
U001-1000-HR & 1 & 0.01 & 1000 & Yes & $200^2$ \\
U001-2000 & 1 & 0.01 & 2000 & Yes & $100^2$ \\
U001-2000-HR & 1 & 0.01 & 2000 &  Yes  & $200^2$ \\
U001-5000 & 1 & 0.01 & 5000 & Yes & $100^2$ \\
U001-5000-HR & 1 & 0.01 & 5000 &  Yes  & $200^2$ \\
U001-10000 & 1 & 0.01 & 10000 & Yes & $100^2$ \\
U001-10000-HR & 1 & 0.01 & 10000 &  Yes  & $200^2$ \\
U001-20000 & 1 & 0.01 & 20000 & Yes & $100^2$ \\
U001-20000-HR & 1 & 0.01 & 20000 &  No  & $200^2$ \\
U001-INF & 1 & 0.01 & $\infty$ & No & $100^2$\\
 \hline
\end{tabular}
\caption{Simulation runs performed. The first column is the name of the run, subsequent columns are the energy in the $B_y$ component, the energy in the $B_x$ and $B_z$ components, the value of $R_H$, whether the simulations converges or not and the resolution of the run. }
\label{TAB1}
\end{table}
\begin{table}[h!]
\centering
\begin{tabular}{|l |c |c |c |c |c|} 
 \hline
 Name & $E_{y}$ & $E_{xz}$ & $R_{H}$ & Conv. & Res.\\ [0.5ex]
 \hline
 C010-20 & 1 & 0.1 & 20 & Yes& $100^2$ \\ 
 C010-50 & 1 & 0.1 & 50 & Yes & $100^2$ \\
 C010-100 & 1 & 0.1 & 100 & Yes & $100^2$ \\
 C010-200 & 1 & 0.1 & 200 & No & $100^2$ \\
 [1ex] 
 \hline
U010-50 & 1 & 0.1 & 50 & Yes & $100^2$ \\ 
U010-100 & 1 & 0.1 & 100 & Yes & $100^2$ \\
U010-200 & 1 & 0.1 & 200 & Yes & $100^2$ \\
U010-500 & 1 & 0.1 & 500 & Yes & $100^2$ \\
U010-500-HR & 1 & 0.1 & 500 & Yes & $200^2$ \\
U010-1000 & 1 & 0.1 & 1000 & Yes & $100^2$ \\
U010-1000-HR & 1 & 0.1 & 1000 & Yes & $200^2$ \\
U010-2000 & 1 & 0.1 & 2000 & Yes & $100^2$ \\
U010-2000-HR & 1 & 0.1 & 2000 &  Yes  & $200^2$ \\
U010-5000 & 1 & 0.1 & 5000 & Yes & $100^2$ \\
U010-5000-HR & 1 & 0.1 & 5000 &  No    & $200^2$ \\
U010-10000 & 1 & 0.1 & 10000 & Yes & $100^2$ \\
U010-20000 & 1 & 0.1 & 20000 & No & $100^2$ \\
 \hline
\end{tabular}
\caption{Simulation runs performed. Columns are the same as in Table \ref{TAB1}.}
\label{TAB2}
\end{table}
\begin{table}[h!]
\centering
\begin{tabular}{|l |c |c |c |c |c|} 
 \hline
 Name & $E_{y}$ & $E_{xz}$ & $R_{H}$ & Conv. & Res.\\ [0.5ex]
 \hline
 C050-20 & 1 & 0.5 & 20 & Yes & $100^2$ \\ 
 C050-50 & 1 & 0.5 & 50 & Yes & $100^2$ \\
 C050-100 & 1 & 0.5 & 100 & Yes & $100^2$ \\
 C050-200 & 1 & 0.5 & 200 & No & $100^2$ \\
 [1ex] 
 \hline
U050-50 & 1 & 0.5 & 50 & Yes & $100^2$ \\ 
U050-100 & 1 & 0.5 & 100 & Yes & $100^2$ \\
U050-200 & 1 & 0.5 & 200 & Yes & $100^2$ \\
U050-500 & 1 & 0.5 & 500 & Yes & $100^2$ \\
U050-500-HR & 1 & 0.5 & 500 & Yes & $200^2$ \\
U050-1000 & 1 & 0.5 & 1000 & No & $100^2$ \\
U050-1000-HR & 1 & 0.5 & 1000 & No & $200^2$ \\
 \hline
\end{tabular}
\caption{Simulation runs performed. Columns are the same as in Table \ref{TAB1}.}
\label{TAB3}
\end{table}
\begin{table}[h!]
\centering
\begin{tabular}{|l |c |c |c |c |c|} 
 \hline
 Name & $E_{y}$ & $E_{xz}$ & $R_{H}$ & Conv. & Res.\\ [0.5ex]
 \hline
 C100-20 & 1 & 1 & 20 & Yes & $100^2$ \\ 
 C100-50 & 1 & 1 & 50 & Yes & $100^2$ \\
 C100-100 & 1 & 1 & 100 & No & $100^2$ \\
 [1ex] 
 \hline
U100-20 & 1 & 1 & 20 & Yes & $100^2$ \\
U100-50 & 1 & 1 & 50 & Yes & $100^2$ \\ 
U100-100 & 1 & 1 & 100 & Yes & $100^2$ \\
U100-200 & 1 & 1 & 200 & Yes & $100^2$ \\
U100-500 & 1 & 1 & 500 & No & $100^2$ \\
U100-1000 & 1 & 1 & 1000 & No & $100^2$ \\
 \hline
\end{tabular}
\caption{Simulation runs performed. Columns are the same as in Table \ref{TAB1}.}
\label{TAB4}
\end{table}

We have performed 54 runs implementing central difference and Upwind scheme, using a variety of combinations of the magnetic field components and $R_H$. Below we analyse the behaviour of the magnetic fields for the following runs. 

In the series of C001 and U001 runs (Table \ref{TAB1}), we notice that $B_y$ is supported by a current on the $x-z$ plane that swirls the $B_x-B_z$ field, Figure \ref{Fig:1}. Even though the planar field is weak it adopts a state where contours of $B_{y}$ and $\Psi$ coincide, especially for high values of $R_{H}$. 

Finite difference schemes are known to produce numerical dissipation, in this particular context it is physically interpreted as numerical resistivity. In the standard resolution runs we performed ($100^2$) we noticed that the Upwind scheme was converging for very high values of $R_H$, especially when $E_{xz}$ was low (0.01 and 0.1). To assess whether this convergence was because of the presence of numerical resistivity we run the simulations at higher resolution that would presumably have lower numerical resistivity and we also run a simulation where we switched-off completely the Ohmic term. We find that the the latter case U001-INF, was diverging, however, runs U001-20000 was converging while runs U001-20000-HR and U001-10000-HR were diverging. This implies that the runs are affected by numerical dissipation at this stage leading to convergence. This effect also occurs in the runs where the energy of the planar component is $10\%$ of the energy of the vertical component, where U010-5000 converges but run U010-5000-HR diverges. It is no longer the case for runs where the planar component energy is $50\%$ of the energy of the vertical component, as there both runs U050-1000 and U050-1000-HR diverge and the maximum convergence is achieved for $R_H=500$ in both resolutions. As these runs were already demanding, with a typical run lasting a few days, we have not explored a higher resolution. 

In runs with higher values of the planar magnetic field, we notice that the evolution becomes more complex as the $B_x-B_z$ field interact with $B_y$. This effect can be seen in the runs of the U010, U050 and U100 in Figures \ref{Fig:2}-\ref{Fig:4}. While initially the vertical field drives the evolution, the planar field interacts with the vertical. Quite interestingly, late in the evolution, the field tends to adopt a particular structure figures \ref{Fig:2}-\ref{Fig:4}, with the planar and vertical fields following similar contours. 

In all families of runs we notice that the simulations employing the Upwind scheme converge for substantially higher values of $R_H$ compared to the central difference runs. In the runs where the energy in the planar field was $0.01$ of the vertical field, central difference simulations could reach a maximum value of $R_H=200$, whereas Upwind runs could reach exceptionally high values of $R_{H}$ exceeding $10^{3}$. We need to stress that the choice of the numerical derivative in the FTCS is second order (being central difference) and is outperformed by the Upwind scheme where the forward or backward difference scheme used is first order.

\begin{figure}
a\includegraphics[width=0.45\textwidth]{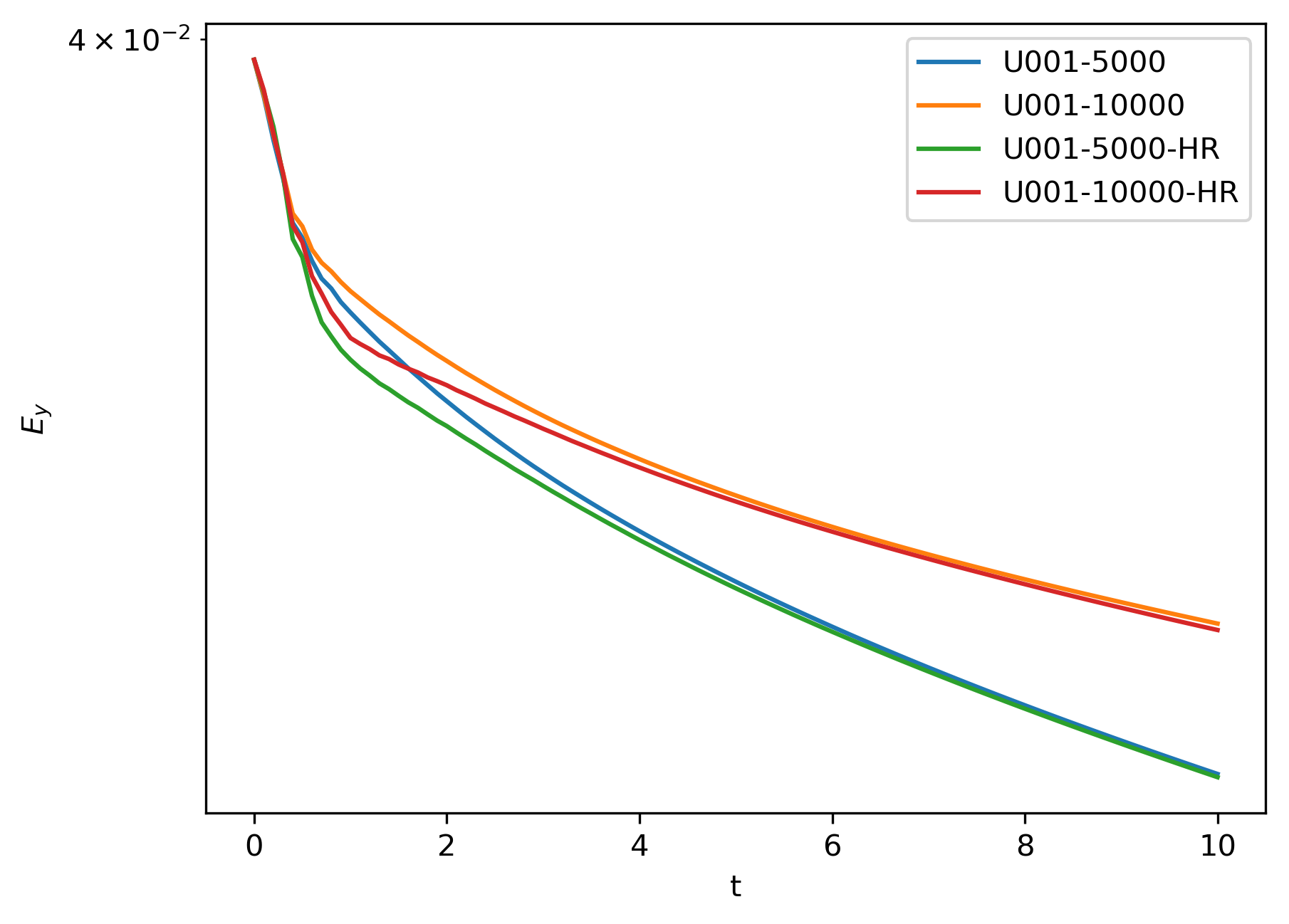}
b\includegraphics[width=0.45\textwidth]{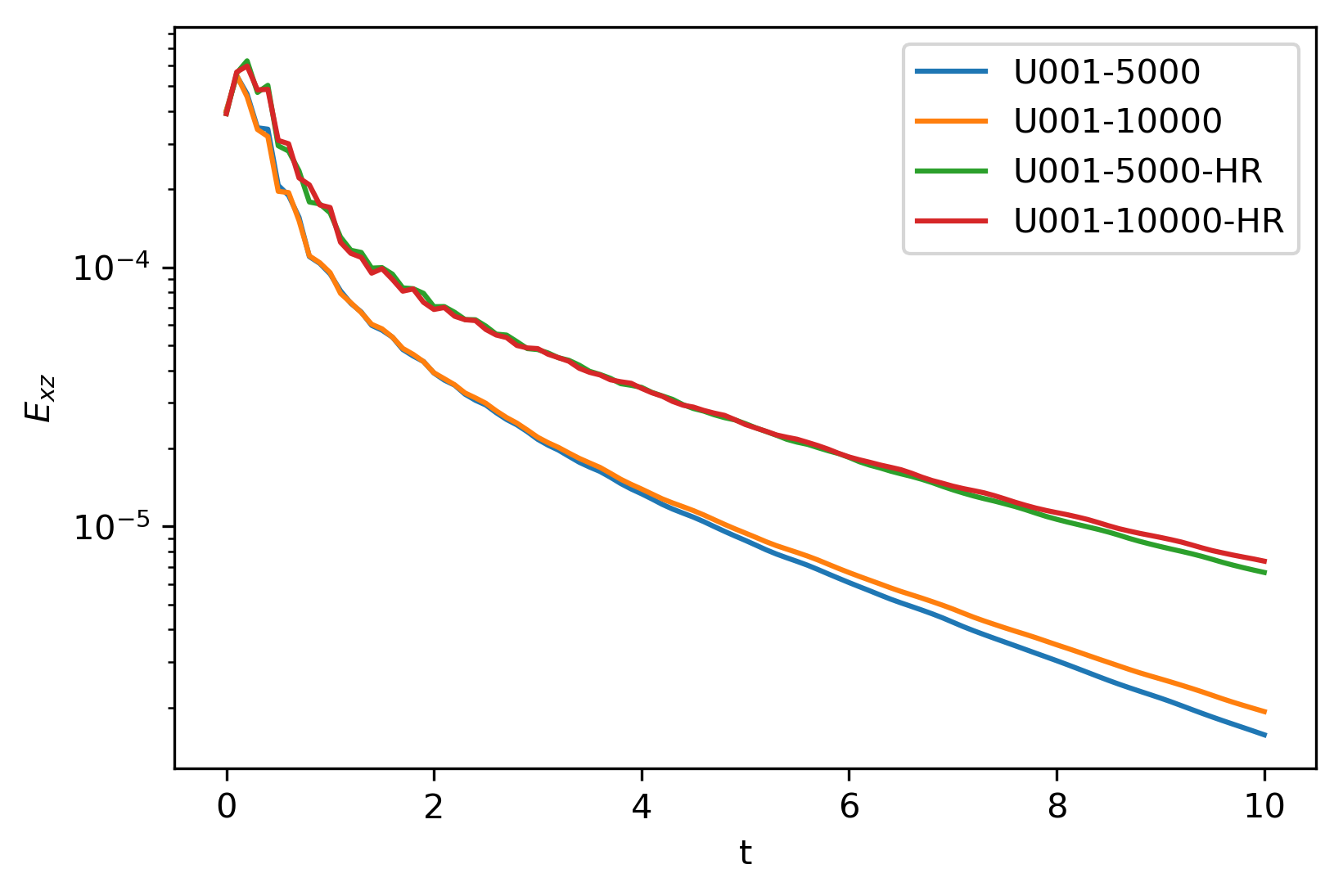}
c\includegraphics[width=0.45\textwidth]{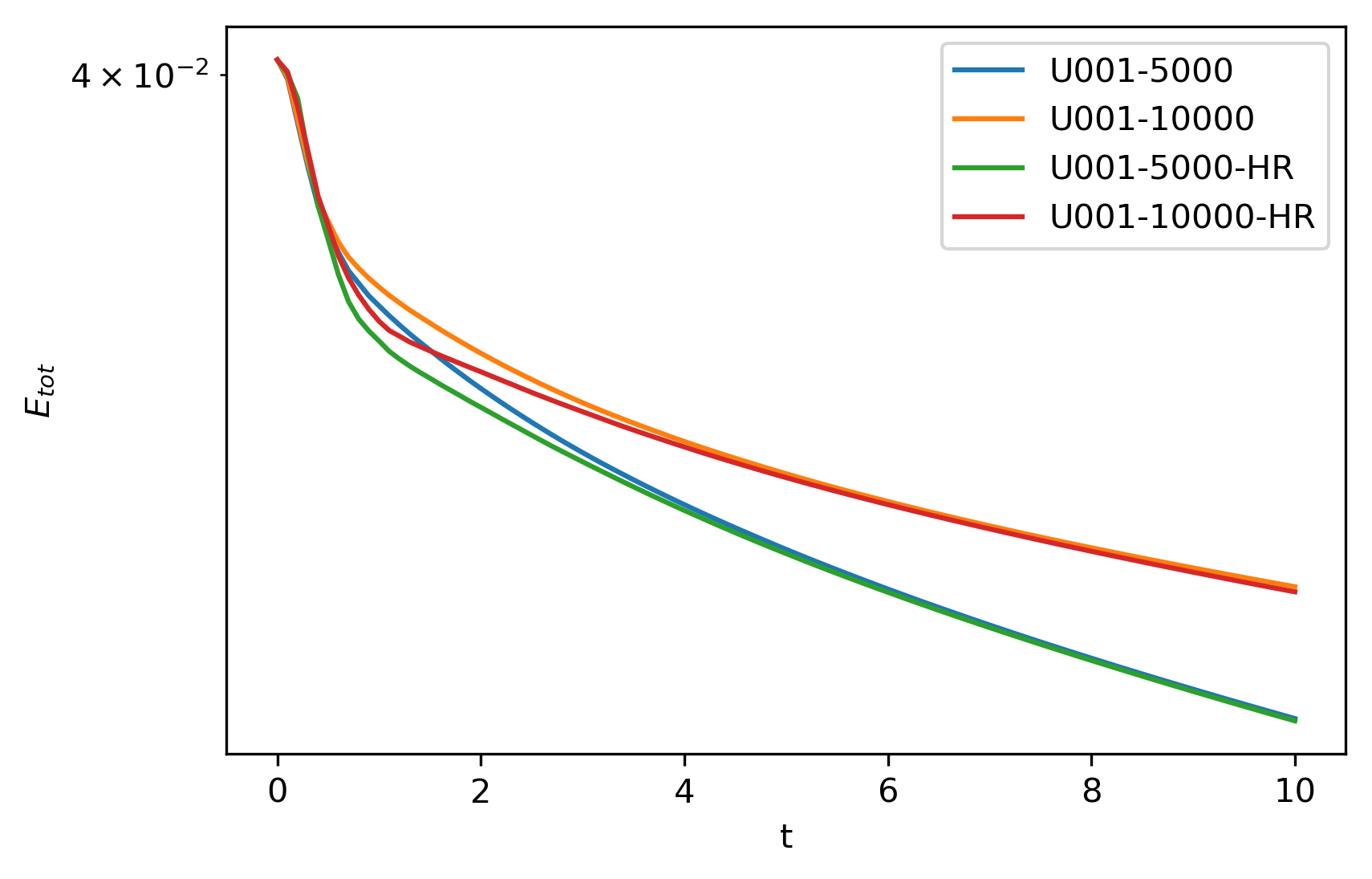}
\caption{Magnetic energy decay for models U001-5000 (blue), U001-10000 (orange), U001-5000-HR (green) and U001-10000-HR (red). In the top panel (a) the energy of the vertical component is plotted, in the middle panel (b) the energy of the planar component and in the bottom panel (c) the total energy. }
\label{Fig:6} 
\end{figure}
\begin{figure}
a\includegraphics[width=0.45\textwidth]{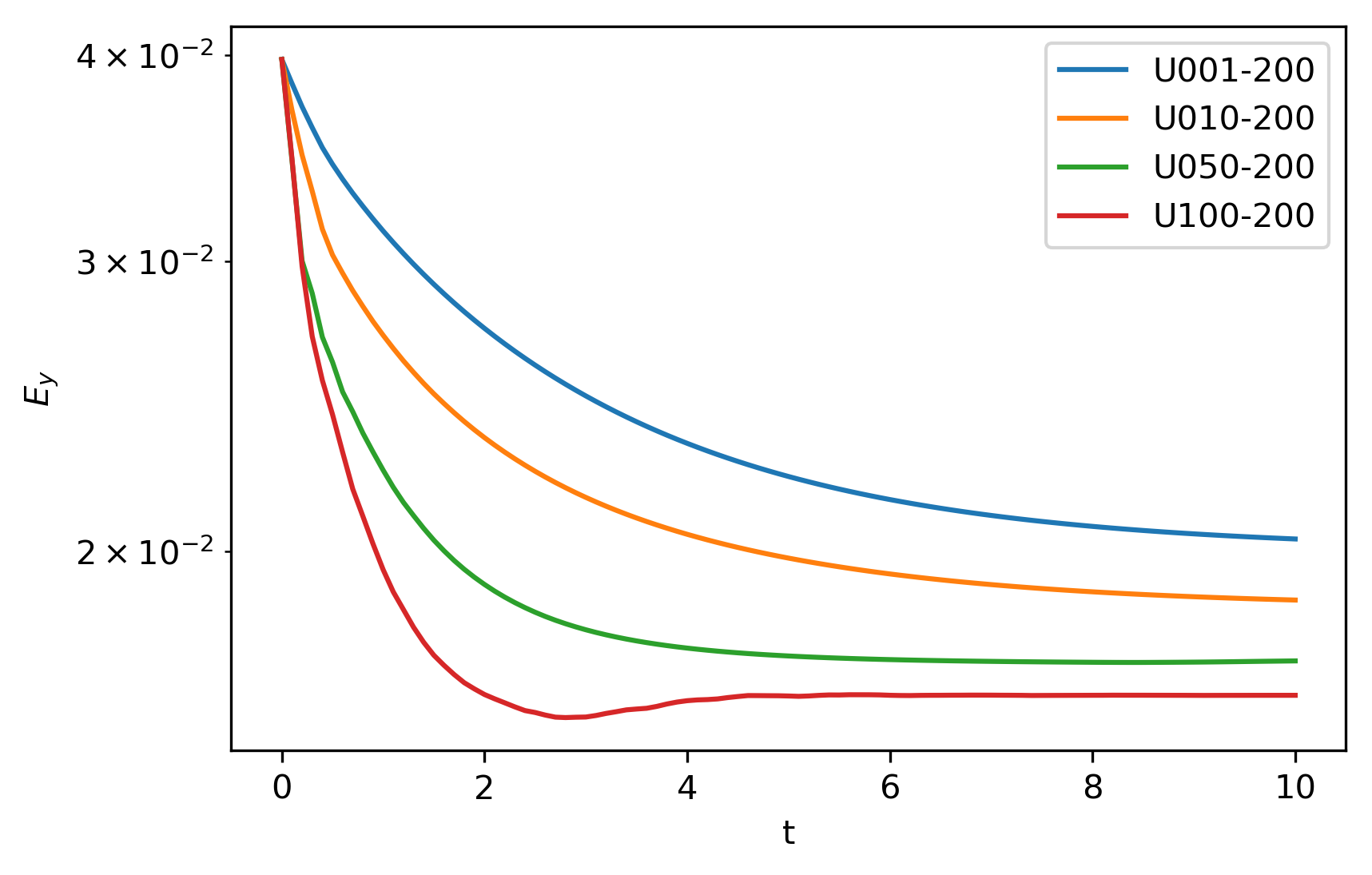}
b\includegraphics[width=0.45\textwidth]{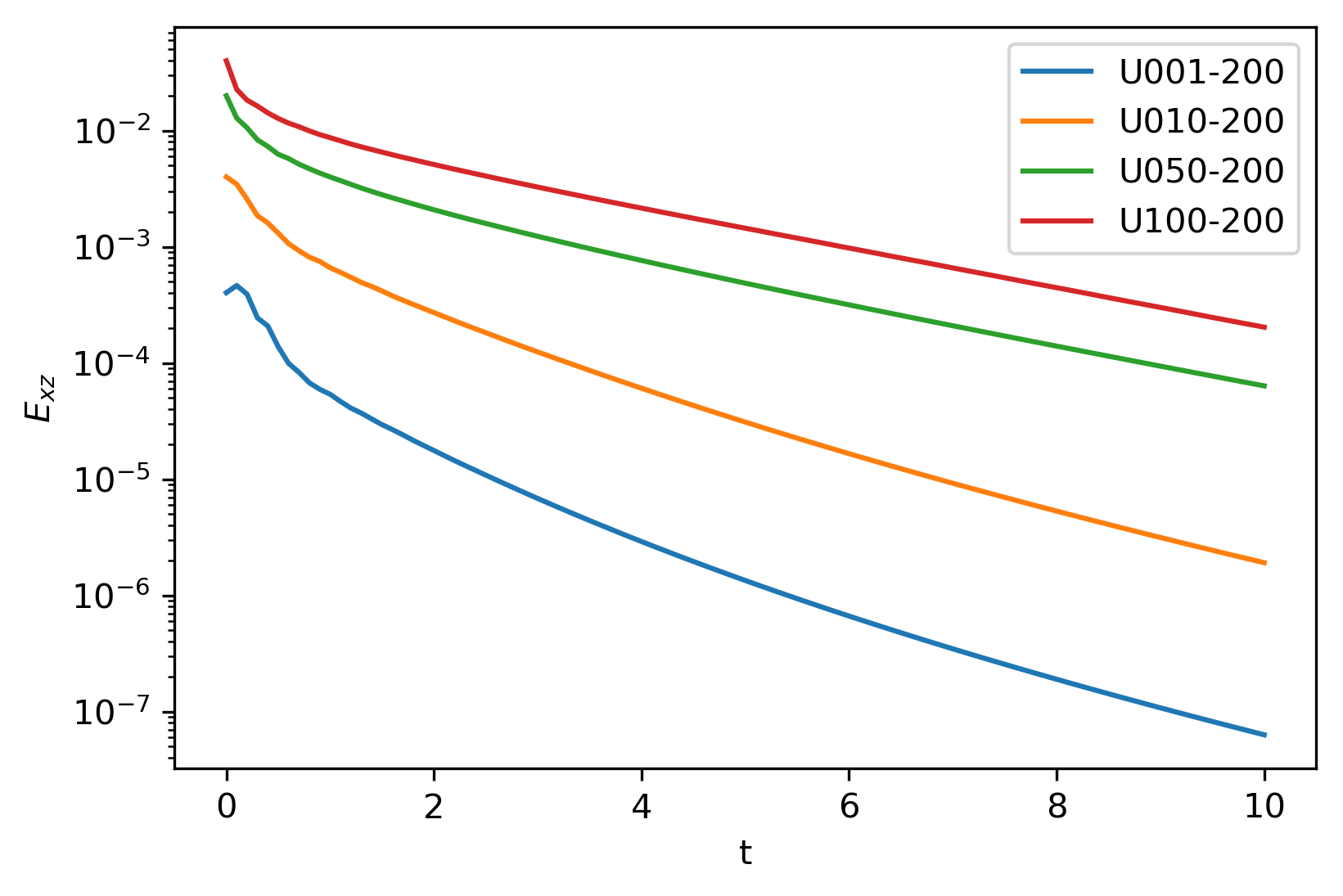}
c\includegraphics[width=0.45\textwidth]{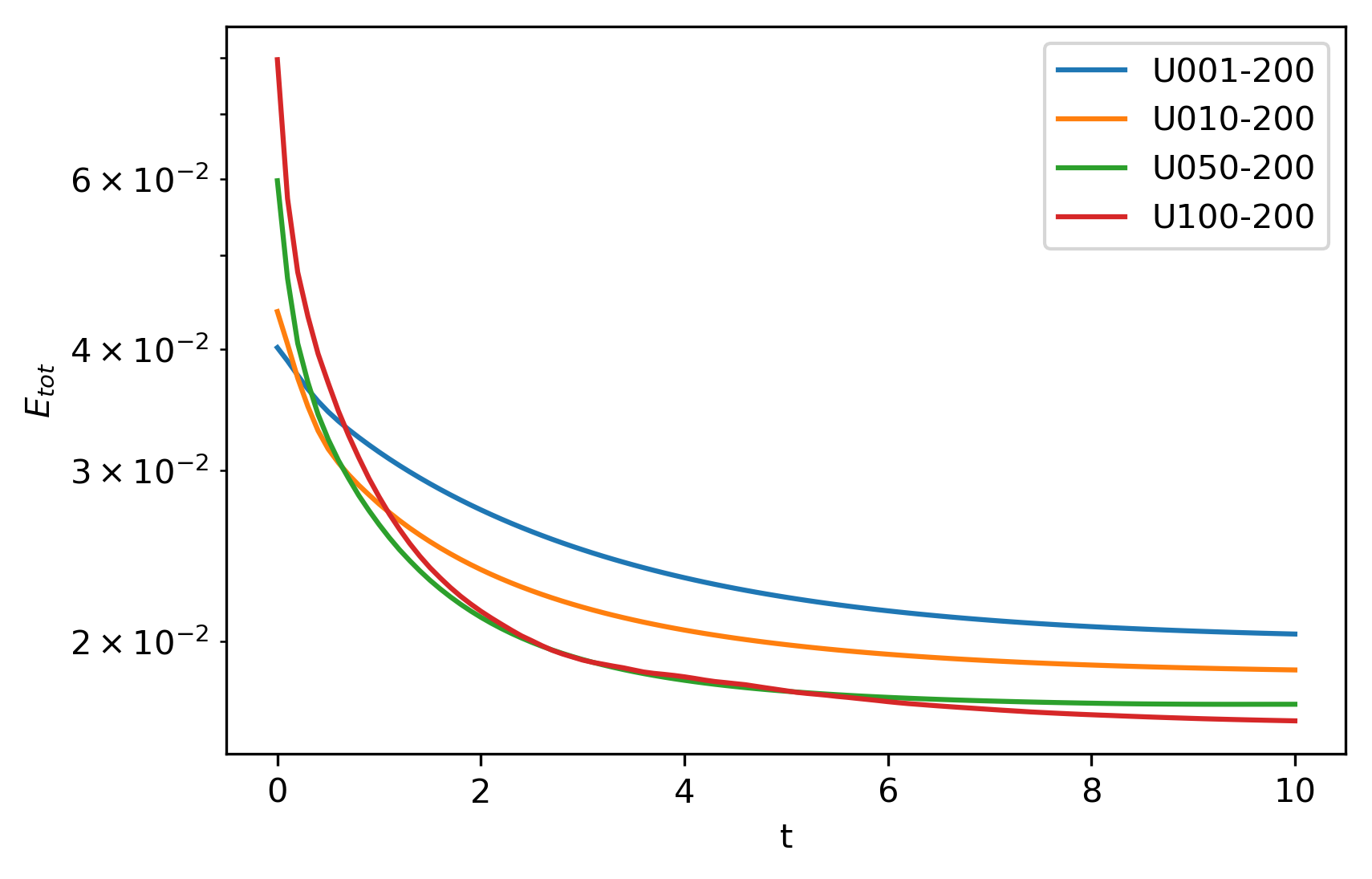}
\caption{Magnetic energy for models U001-200 (red), U010-200 (green), U050-200 (orange) and U100-200 (blue). In the top panel (a) the energy of the vertical component is plotted, in the middle panel (b) the energy of the planar component and in the bottom panel (c) the total energy. }
\label{Fig:7} 
\end{figure}
In the families with progressively higher values of planar field (C010 and U010; C050 and U050; C100 and U100) we notice that in both FTCS and Upwind runs the maximum value of $R_H$ for convergence decreases, but the Upwind scheme still converges for values of $R_H$ that are higher by an order of magnitude or at least a factor of a few. For instance in the runs where the energy on the planar and vertical field are equal (C100 and U100) FTCS scheme converges for up to $R_H=50$ whereas the Upwind scheme converges for $R_H=200$. What is further noteworthy, is the way the two schemes diverge. FTCS always diverge through a very small time-step due to the Courant condition imposed, while the energy diverges exponentially. On the contrary, in the runs U050-1000 and U100-500 we notice that temporarily the energy increases instead of decreasing. Despite that, the time-step does not become zero and still the code continues to run, eventually dissipating any numerical instabilities. Even in the runs that converged for both the FTCS and Upwind schemes, i.e.~U001-200 and C001-200, there are some noticeable differences. Runs employing FTCS tend to create planar magnetic fields whose lines of force have sharper angles than the ones using the Upwind scheme which are smoother, Figure \ref{Fig:5}. Because of this, they are more prone to numerical instability. 

The fact that the improvement of the code using the Upwind scheme is not so dramatic when $\Psi$ becomes large is due to the fact that the Upwind scheme operates on the term containing the advective velocity appearing in the form of derivatives of $B_y$. Clearly, in the aforementioned runs, the significance of the first term in the right-hand-side of equation \ref{dBy} becomes higher and affects the overall convergence of the run. Thus, even if the use of the Upwind scheme is a major improvement in the first term of the right-hand-side of equation \ref{dPsi}, it still can not secure the simulation from instabilities that arise from non-advective terms, such as this term bending the planar field into the vertical one.  

The energy decay in models U001-5000 and U001-10000-HR both in high and low resolution is the same for the both resolutions in the long run, Figure \ref{Fig:6}. However, in the first stages of evolution the resolution plays a significant role, suggesting that the effects observed are more affected by the resolution rather than the resistivity. Regrading the planar component, that is energetically subdominant, it shows initially some oscillations, due to the effect of swirling from the vertical field. These effects are affected by the resolution of the scheme. This suggests that a choice of $R_{H}$ in the order of $10^3$, even if it leads to a scheme that converges numerically, it may  concealing some of the finer features of the evolution which have some non-trivial impact. 

Comparing runs with different initial ratios of planar to vertical fields, Figure \ref{Fig:7}, we notice, that the runs with stronger planar fields have an overall faster decay, due to the interplay of the planar with the vertical field and the more complex evolution. As we have chosen here a value of $R_{H}=200$, we notice that the planar field eventually decays exponentially, while the vertical converges to a constant value, that of a uniform field. This is evident also from Figures \ref{Fig:3} and \ref{Fig:4} panel (d) where in both cases the vertical field becomes uniform, whereas the planar adopts a structure of a single maximum. This is related to our choice of the initial topology of the field, with $B_{y}$ having a positive net flux, whereas $\Psi$ having a single maximum.

\section{Conclusions}
\label{Con}

In this paper we have shown that the use of an Upwind scheme greatly improves the convergence of Hall-MHD numerical simulations. The scheme is applied in the advective term and decreases the possibility of the appearance of numerical instability, allowing for simulations with higher values of $R_H$ compared to runs using the FTCS scheme. The improvement can be approximately two orders of magnitude for runs where the planar magnetic field has $1\%$ of the energy compared to the vertical, and a factor of a few for the runs where the two components contain the same amount of energy. The constraint of a maximum value in $R_H\sim 100$ for the FTCS scheme is in agreement with previous works that employed a plane-parallel geometry \citep{2015MNRAS.453L..93G, 2016MNRAS.463.3381G}. 

This improvement can enhance the efficiency of runs of Hall-MHD simulations allowing for higher values of conductivity, which is closer to realistic conditions for magnetars. Because of such constraints, most models employ crusts where the Hall parameter is lower than what would realistically be expected, mostly reflecting a numerical hurdle rather than a physical limitation. We note that in the present work, we have explored a plane-parallel model of a crust with uniform density, where the Upwind scheme was only applied to the advection term of the flux $\Psi$. In the more realistic approach of a stratified density, there is an additional advection term in the $B_y$ component corresponding to a non-linear shock wave similar to Burger's Equation. This term could be further implemented with an Upwind scheme. Even in a uniform density crust, subject to the constraint of axisymmetry, there exists an advective term, which needs to be accounted for in the calculation. Things may become more complex in the state-of-the art three-dimensional simulations \citep{2015PhRvL.114s1101W}. These simulations are spectral and the implementation of an Upwind scheme would require a different framework for its operation, possibly using a finite difference approach. This type of approach could however introduce a variety of other problems such as slowing of the numerical code and singularities on the axis. 

Overall we conclude that the inclusion of an Upwind scheme, even if it is technically simple to implement, it leads to an impressive improvement of the performance of the numerical code, and is worth exploring its application in different contexts such as axisymmetric simulations.

%\clearpage
%\newpage
%%%%%%%%%%%%%%%%%%%%%
%%%%%%%%%%%%%%%%%%%%%
%%%%%%%%%%%%%%%%%%%%%

\section{Acknowledgements}
The authors are grateful to E.P.Christopoulou, N.Vlahakis and V. Karageorgopoulos for numerous englightening discussions and comments.

\printcredits

\bibliographystyle{elsarticle-harv}

\bibliography{cas-refs}
%\bibliography{elsarticle-harv}

\end{document}